\documentstyle[12pt,epsf,eqsecnum,aps,amssymb,amsfonts]{revtex}

\newlength{\ul}
\begin{document}

\sloppy

\draft

\title{Critical Casimir Forces in Colloidal Suspensions}
\author{\\F. Schlesener,$^{1,2}$ A. Hanke,$^{3,4}$ and S. Dietrich$^{1,2}$}
\address{$^1$Max-Planck-Institut f\"ur Metallforschung,
Heisenbergstr. 1, D-70569 Stuttgart, Germany}
\address{$^2$Institut f\"ur Theoretische und Angewandte Physik,
Universit\"at Stuttgart,\\
D-70550 Stuttgart, Germany}
\address{$^3$Department of Physics, Massachusetts Institute of Technology,
Cambridge, MA 02139, USA}
\address{$^4$Department of Physics, Theoretical Physics, 1 Keble Road,
Oxford OX1 3NP, United Kingdom}
\date{\today}
\maketitle

\tightenlines

\begin{abstract}
Some time ago, Fisher and de Gennes pointed out that
long-ranged correlations in a fluid close to its critical point $T_c$ cause 
distinct forces between immersed colloidal particles which can even lead to
flocculation [C. R. Acad. Sc. Paris B {\bf 287}, 207 (1978)].
Here we calculate such forces between pairs of spherical particles
as function of {\em both\/} relevant thermodynamic variables, i.e., 
the reduced 
temperature $t = (T - T_c) / T_c$ and the field $h$ conjugate to the order 
parameter. This provides the basis for specific predictions 
concerning the phase behavior 
of a suspension of colloidal particles in a near-critical solvent.
\end{abstract}

\bigskip
\pacs{Keywords: Critical phenomena, Casimir forces, colloids.}

%%%%%%%%%%%%%%%%%%%%%%%%%%%%%%%%%%%%%%%%%%%%%%%%%%%%%

\section{Introduction}
\label{intro}

In 1978 Fisher and de Gennes \cite{FdG78} predicted that 
the confinement of critical fluctuations of the order parameter
in a binary liquid mixture near its critical demixing point 
$T_c$ gives rise to long-ranged forces between immersed plates or
particles. In particular, they pointed out that these long-ranged 
forces would eventually lead to the flocculation
of colloidal particles which are dissolved in a near-critical 
binary liquid mixture \cite{dg81}.
A few years later, such a solvent-mediated flocculation
was observed experimentally for silica spheres immersed in 
the binary liquid mixture of water and 2,6-lutidine \cite{BE,GKM92}. 
However, the precise interpretation of these experiments is still 
under debate; although fluctuation-induced forces as predicted 
by Fisher and de Gennes certainly play a major role, additional 
mechanisms, such as screening effects in the case of charged colloidal 
particles \cite{BE}, may also contribute \cite{LPB98}.
In the meantime additional experimental evidence for this kind of
flocculation phenomena has emerged for other binary mixtures acting
as solvents \cite{NK93,JK97,GW97}.

From a colloid physics perspective effective, solvent-mediated
interactions between dissolved colloidal particles are of basic 
importance \cite{dispersions,colloid,SHKD2001,likos}. The richness of the
physical properties of these systems is mainly based on the 
possibility to tune these effective interactions over wide
ranges of strength and form of the interaction potential.
Traditionally, this tuning is accomplished by changing the 
chemical composition of the solvent by adding salt, polymers, 
or other components \cite{dispersions,colloid,SHKD2001,likos}. 
Compared with such modifications, changes of the temperature 
or pressure typically result only in minor changes of the 
effective interactions.
On the other hand, however, effective interactions generated by bringing
solvents close to a phase transition of their own are extremely
sensitive to such changes. In this broader context, flocculation
of colloidal particles induced by critical fluctuations near a
{\em second-order\/} phase transition of the solvent
as predicted by Fisher and de Gennes \cite{FdG78,dg81}
provided the first instance in which effective interactions are 
tuned by changing the physical conditions rather than the chemical 
composition of the solvent. Later on, this idea has been extended 
to {\em first-order\/} transitions of the solvent, for which
wetting transitions can occur which give rise to wetting films of 
the preferred phase coating the colloidal particles 
\cite{lowen,BD98,BBD2000,BM2001}. In the meantime the study of effective
interactions between dissolved particles generated by changing the physical 
conditions of the solvent, and the resulting phase behavior of the 
colloidal suspension, has emerged as one of the main fields of research 
in colloid physics \cite{BD98}.

The physical origin of the force predicted by Fisher and de Gennes
is analogous to the Casimir force between conducting plates, 
which arises due to the confinement of quantum mechanical vacuum 
fluctuations of the electromagnetic field \cite{C48,Lam99,BCOR02}.
(Related phenomena occur in string theory \cite{S81}, liquid
crystals (see, e.g., ref. {}\cite{ZPZ99}), and microemulsions
\cite{U01}. However, in contrast to the present so-called ``critical Casimir effect'', in liquid crystals the director-fluctuation 
induced forces are notoriously difficult to separate from
the background of dispersion forces due to the absence of a singular 
temperature dependence \cite{ZM01}.)
In recent years this ``critical Casimir effect'' has attracted 
increasing theoretical \cite{INW86,K91,Krech,K97,KG99} and experimental
\cite{LPB98,ML99,GC} interest.
In this case the fluctuations are provided by the critical order parameter
fluctuations of a bona fide second-order phase transition in the
corresponding bulk sample.
For He$^4$ wetting films close to the superfluid phase transition the
theoretical predictions \cite{K91} for critical Casimir forces
between parallel surfaces exhibiting
Dirichlet boundary conditions have been confirmed quantitatively \cite{GC}.
These quantum fluids offer the opportunity to study critical
Casimir forces in the absence of surface fields, i.e., the purely
fluctuation induced forces.

In a classical binary liquid mixture near its critical demixing point,
the order parameter is a suitable 
concentration difference between the two species forming the liquid.
In this case the
inevitable preference of confining boundaries for one of the two species
results in the presence of effective surface fields leading to
nonvanishing order parameter profiles even at $T \ge T_c$ \cite{fluid}.
(This is the case discussed by Fisher and de Gennes \cite{FdG78,dg81}.)
These so-called ``critical adsorption'' profiles become 
particularly long-ranged due to the correlation effects induced by 
the critical fluctuations of the order parameter of the solvent. 
In the case of a planar wall, critical adsorption has been studied
in much detail \cite{B83,D86,LF89,C90,DS93,ES94,FD95,Law2001}. 
Asymptotically close to the critical point $T_c$
it is characterized by a surface field which is infinitely 
large so that the order parameter profile actually diverges 
upon approaching the wall up to atomic distances.
Later on, critical adsorption on curved surfaces
of {\em single\/} spherical and rodlike colloidal particles has 
been studied, where it turns out, in particular, that 
critical adsorption on a microscopically thin ``needle'' 
represents a distinct universality class of its own 
\cite{HD99,Han2000}. Critical adsorption on the rough interface
between the critical fluid and its noncritical vapor occurring
at the critical end point of the binary liquid mixture has been
addressed as well \cite{DS}; likewise the effect of quenched 
surface modulations and roughness \cite{HK2001}.
The interference of critical adsorption 
on {\em neighboring spheres\/} gives rise to the critical Casimir forces
which have been argued to contribute to the occurrence of flocculation
near $T_c$ \cite{dg81,BE95,N96,HSED98}. However, a quantitative 
understanding of these phenomena requires 
the knowledge of the critical adsorption profiles near the colloidal 
particles and the resulting effective free energy of interaction 
in the whole vicinity of the critical point, i.e., as
function of both the reduced temperature $t = (T - T_c) / T_c$ and 
the bulk field $h$ conjugate to the order parameter. So far, this 
ambitious goal has not yet been accomplished. Instead, 
the introduction of the additional complication of a surface curvature 
has limited the knowledge of the corresponding critical adsorption so 
far to the case of spheres for the particular 
thermodynamic state $(t,h) = (0,0)$ of the solvent
\cite{BE95,BE85,GR95,ER95}. 

In a previous work \cite{HSED98}, at least the temperature dependence 
of the critical Casimir force between a sphere and a planar container 
wall has been discussed. It turns out
that the force becomes maximal not at but {\em above\/}
$T_{c\,}$. In ref. {}\cite{HSED98} also the force $F$ 
between a pair of spheres of radius $R$ at distance $D+2\,R$
between their centers (see Fig.\,\ref{fig_geo}) has 
been briefly analyzed. 
%
% Fig.1
%
Close to the critical demixing 
point, this force separates into a regular background contribution
and a singular contribution $F^{\pm}_{\text{sing}}$ of universal
character, which is {\em attractive\/} if the same of the 
two coexisting bulk phases is enriched near both spheres. 
The results for $F^{\pm}_{\text{sing}}$ can be cast into the form
\begin{equation}
\label{scalf}
F^{\pm}_{\text{sing}}(D,R;t,h) \, = \, \frac{k_B T_c}{R} \,
K_{\pm}\left(\Delta = \frac{D}{R}\, , \, \Theta_{\pm}=\frac{D}{\xi_{\pm}}\,,\,
\Xi = \text{sgn}(h)\,\frac{D}{l_H} \right) \, \, ,
\end{equation} 
where $\xi = \xi_{\pm}$ is the {\em true} correlation length 
governing the exponential decay of the order parameter correlation
function in the bulk for $t = (T - T_c)/T_c \gtrless 0$ at $h = 0$;
$l_H$ is the correlation length 
governing the exponential decay of the order parameter correlation
function for $h \to 0$ at $t = 0$. $K_{\pm}$ are universal scaling functions.
Note that surface fields $h_1$ break the symmetry $h\,\to\,-h$, i.e.,
$F^{\pm}_{\text{sing}}$ depends on the sign of $\Xi$.
Here and in the following we consider only the fixed point value 
$h_1\to +\infty$.

The aim of this work is to study the critical Casimir force
between two spherical colloidal particles as function of {\em both\/} 
relevant thermodynamic variables $t$ and $h$. 
Figure$\,$\ref{fig_pdp} shows the schematic phase diagram of the solvent,
e.g., a binary liquid mixture. Different thermodynamic paths are
indicated along which the dependence of the force on $t$ and $h$ is studied.
%
% Fig.2
%

The remainder of this work is organized as follows:
In Sec.\,\ref{sec_mod} we define the field-theoretic model used 
to describe the near-critical solvent.
In Sec.\,\ref{sec_num} the numerical calculation 
of the adsorption profiles in mean-field approximation is discussed.
In Sec.\,\ref{sec_ses} we present the results for the force
between two spherical particles and compare them with the 
limiting cases of small and large particle radii. 
 The discussion 
of the results for two parallel plates, which inter alia are relevant for a
pair of spheres via the Derjaguin approximation, is left to 
Sec.\,\ref{sec_flm}.
Finally, in Sec.\,\ref{sec_con}, we summarize and discuss our
results with regards to the phase behavior of a suspension 
of colloidal particles in a near-critical fluid.

%%%%%%%%%%%%%%%%%%%%%%%%%%%%%%%%%%%%%%%%%%%%%%%%%%%%%%%%%%%%%%%%%%

\section{Field-theoretic model}
\label{sec_mod}
Near criticality the behavior of the system is governed by fluctuations
of the order parameter $\Phi$
on a large length scale such that the emerging universal properties
are independent of microscopic details. A coarse-graining procedure
leads to the continuum description in terms of
the Ginzburg-Landau Hamiltonian \cite{B83,D86,F74}
\begin{equation} \label{I100}
{\cal H}\{ {\Phi}({\bf r}) \} =  
\int\limits_{V}
d^d r \, \Big\{ \frac{1}{2} [\nabla {\Phi}({\bf r})]^2
\, + \, \frac{\tau}{2} \, {\Phi}^2({\bf r})
\, + \, \frac{u}{24} \, {\Phi}^4({\bf r}) \, - \, 
{h} \, {\Phi}({\bf r}) \Big\} \,
\end{equation}
of an order parameter field ${\Phi}({\bf r})$ in the bulk of volume $V$.
For the liquids considered here ${\Phi}({\bf r})$ is a scalar field.
The Hamiltonian determines the statistical weight 
$\exp[-{\cal H}\{{\Phi}({\bf r})\}]$ of the configuration
$\{{\Phi}({\bf r})\}$. 
The external bulk field ${h}({\bf r})$ is conjugate
to the order parameter and breaks its symmetry.
The temperature dependence enters via
\begin{equation}
\label{If1}
\tau\sim t=\frac{T-T_c}{T_c}\quad.
\end{equation}
The partition function of the system is given by

\begin{equation}
\label{If2}
{\cal Z}(\tau, h; u)=\int{\cal D}\{{\Phi}\}
\,\exp[-{\cal H}\{{\Phi}({\bf r})\}]\quad,
\end{equation}
where ${\cal D}\{{\Phi}\}$ denotes an appropiately defined
functional integration \cite{D86}.
The corresponding singular part of the free energy reads 
\begin{equation}
\label{fsing}
{\cal F}_{\rm sing}(\tau, h; u)=-k_B\,T_c\,\ln{\cal Z}(\tau, h; u)\;.
\end{equation}

The coarse-graining procedure leading to a continuum model 
from a given lattice model can be generalized to systems with 
surfaces resulting in the Hamiltonian \cite{B83,D86}
\begin{eqnarray}
{\cal H}\{ {\Phi}({\bf r}) \} \, & = & \,  
\int\limits_{V}
d^d r \, \Big\{ \frac{1}{2} [\nabla {\Phi}({\bf r})]^2
\, + \, \frac{\tau}{2} \, {\Phi}^2({\bf r})
\, + \, \frac{u}{24} \, {\Phi}^4({\bf r}) \, - \, 
{h} \, {\Phi}({\bf r}) \Big\} \nonumber\\[2mm]
& + & \, \int\limits_{S}
dS \, \Big\{ 
\frac{c}{2} \, {\Phi}^2({\bf r}_{S}) \, - \, 
{h}_1 \, {\Phi}({\bf r}_{S}) \Big\} \;.\label{I130}
\end{eqnarray}
The first integral runs over the volume $V$ available
for the critical medium, which is bounded by the surface(s) $S=\partial\,V$.
Apart from the restricted space integration, the first integral in
Eq.\,(\ref{I130}) has the same form as the Hamiltonian
(\ref{I100}) for bulk systems disregarding surfaces. The second 
integral in Eq.\,(\ref{I130}) represents a surface contribution 
in which the integral runs 
over the boundary surface $S=\partial\,V$; $c$ is related
to the strength of the
coupling of critical degrees of freedom near the surface
and $h_1$ is the surface analogue of the bulk field $h$ \cite{B83,D86}.

The systematic treatment of the critical fluctuations as 
implied by Eq.$\:$(\ref{If2})
and a justification of the 
form of ${\cal H}\{{\Phi}({\bf r}) \}$
is provided by the field-theoretic renormalization group
framework \cite{D86}. The basic idea is to consider the 
physical system on larger and larger length scales and to 
single out those quantities which eventually remain unchanged 
under the repeated action of such scale transformations.
Such quantities tend
to their {\em fixed point\/} values right at the critical 
point \cite{F74}. For bulk systems, 
there remain only two relevant adjustable parameters,
corresponding to $\tau$ and $h$.
The surface
generates a subdivision of a given bulk universality 
class into different {\em surface universality classes\/}
characterized by different fixed point values for $c$ and $h_1$. 
For $h_1 = 0$ the corresponding leading singular 
behaviors are determined by 
the fixed-point values $c = + \infty$ for the 
ordinary transition (O), $c = 0$ for the special or surface-bulk transition,
and $c = - \infty$ for the extraordinary or normal 
transition (E) \cite{B83,D86}.
For the O and E transitions only $\tau$, $h$, and $h_1$ 
remain as relevant variables. The three cases above were
originally studied for magnetic systems. For liquids,
however, E is the generic case and therefore it is called the {\em normal}
transition. The case $h_1=\infty$ 
with $c$ arbitrary corresponds to critical adsorption, but the
leading singular behavior is the same as for the E
transition \cite{DS93}.

Neglecting fluctuations the Ginzburg-Landau Hamiltonian 
${\cal H}\{ \Phi({\bf r}) \}$ itself represents
a free energy functional which upon 
minimization with respect to $\Phi({\bf r})$ yields
the Euler-Lagrange equations for order parameter profiles
and correlation functions. 
The minimization is equivalent to the {\em mean-field\/}
description of the system,
\begin{equation}
\label{If3}
\left.\frac{\delta\,{\cal H}\{ {\Phi}({\bf r}) \}}
{\delta\,{\Phi}({\bf r})}\right|_{{\Phi}=\langle{\Phi}\rangle}=0\;,
\end{equation}
where the fluctuating order parameter ${\Phi}$ is replaced by its
mean value $\langle{\Phi}\rangle$.
The universal critical behavior 
is captured correctly by mean-field theory 
if the space dimension $d$ exceeds the
upper critical dimension $d_{\text{uc}}=4$.

Because of the relative complexity of the geometry considered here 
(see Fig.$\:$\ref{fig_geo}),
it is not possible to carry out the complete renormalization 
procedure in closed form. Instead, available knowledge 
for bulk systems in $d = 3$
is used in combination with the mean-field solution for the 
geometry considered here in order to estimate universal quantities in $d=3$.
An alternative way to overcome mean-field theory is followed within
the concept of the Local Functional Theory developed in refs.
{}\cite{FU90a,FU90b,BU98,BU01}, in which the free energy is
described by a functional depending {\em locally} on the order parameter
and its derivative. With mean-field theory as a special case,
it takes into account available knowledge in $d=3$, such as the correct 
values for the critical exponents.

\section{Order parameter profiles}
\label{sec_num}
In this Section we describe some technical aspects of the tools and
numerical methods used for obtaining the mean-field results, i.e.,
the determination of the order parameter profiles at the critical point,
the short distance expansion of the profile close to a surface,
the minimization procedure to determine the profiles away from the critical
point, and the calculation of the force between the particles
based on the stress tensor.

Minimization in Eq.$\,$(\ref{If3}) leads
to the differential equation
\begin{equation}
\label{Ib1}
-\Delta\,m+\tau\,m+m^3-H = 0\;,
\end{equation}
where the coupling constant $u$ is absorbed in the order parameter and
in the bulk magnetic field,
\begin{equation}
\label{Ib2}
m= \sqrt{\frac{u}{6}}\,\langle{\Phi}\rangle\:,\;H= \sqrt{\frac{u}{6}}\,h\;.
\end{equation}
Note that we are interested in an estimate 
of the order parameter profile $m$ in $d=3$
where, at the fixed point, $u$ is a positive number.
For $d\nearrow 4$ the
mean-field solution is exact, but the fixed point value of the
coupling constant $u$ vanishes in this limit.

The surface integrals in Eq. (\ref{I130}) lead to a boundary
condition for the order parameter at the surfaces of the
spherical particles. At the critical adsorption fixed point
this boundary condition is given by
\begin{equation}
\label{Ib4}
\Phi_S=\infty\;.
\end{equation}

\subsection{Order parameter profiles at criticality}
\label{sub_con}
While our main interest is the 
dependence of the Casimir force on $t$ and $h$, the knowledge of the order
parameter profile {\em at} the critical point, i.e., for $(t, h)=(0,0)$,
is a useful starting point. In ref. {}\cite{GR95} this latter
profile is given for a critical system confined to a spherical shell
between two concentric spheres with radii $R_{\pm}$ at 
which the order parameter diverges. In terms of elliptic functions it reads
\begin{equation}
\label{Ib5}
m(r)=\frac{\displaystyle\sqrt{\frac{1-k^2}{k^2-\frac{1}{2}}}}
{\displaystyle r\,\mbox{cn}
\left(\sqrt{\frac{1}{2k^2-1}}\:\ln\left(\frac{r}{R_0}\right)\right)}\;,
\end{equation}
where $r$ is the radial distance from the common center of the spheres and
$k$ is the module of the elliptic function $\mbox{cn}$ \cite{GR65}.
The module $k$ must be adjusted such that the profile diverges at
the given radii $R_{\pm}$ leading to the following implicit equation for
$k$:
\begin{equation}
\label{Ib6}
\frac{R_{\pm}}{R_0}=\left(\frac{R_+}{R_-}\right)^{\pm\frac{1}{2}}
=\,\exp{\left(\pm{\sqrt{2k^2-1}}\,{K(k)}\right)}\;,
\end{equation}
where $R_0=\sqrt{R_+\,R_-}$ is the geometric mean of the radii
and $K$ is the complete elliptic integral of the first kind.

In ref. {}\cite{BE95} it is outlined how the results for
concentric spheres can be used for different geometries
via a conformal transformation of the coordinates,
\begin{equation}
\label{Ib7}
\frac{{\bf r}^{\,\prime}}{|{\bf r}^{\,\prime}|^2}=
\frac{{\bf r}+{\bf R}}{|{\bf r}+{\bf R}|^2}
-\frac{{\bf R}}{2\,|{\bf R}|^2}\;,
\end{equation}
where ${\bf r}^{\,\prime}$ and ${\bf r}$ are the new and
the old coordinates, respectively, and ${\bf R}$
defines the symmetry axis and the eventual shape of the new geometry.
For $|{\bf R}|=R_0$ the two concentric spheres with radii $R_{\pm}$ are
mapped onto two spheres of equal size which is the generic case we are
interested in. For $|{\bf R}|=R_+$ the two concentric spheres
are mapped onto a single spherical particle in front of a planar
wall for which results were obtained in ref. {}\cite{HSED98}.

Under the conformal transformation (\ref{Ib7}), each 
local scaling field
is multiplied by a scale factor
\begin{equation}
\label{Ib8}
b({\bf r}^{\,\prime})=1+\frac{{\bf R}\cdot{\bf r}^{\,\prime}}{|{\bf R}|^2}
+\frac{|{\bf r}^{\,\prime}|^2}{4\,|{\bf R}|^2}\;,
\end{equation} 
which is raised to the power of the scaling dimension of this scaling field.
In mean-field approximation and $d = 4$,
the scaling dimension of the order parameter
$\Phi$ is $1$, which implies for the order parameter profile
\begin{equation}
\label{Ib9}
m({\bf r}^{\,\prime})=b({\bf r}^{\,\prime})^{-1}
\,m_{\text{conc}}(|{\bf r}({\bf r}^{\,\prime})|)\;,
\end{equation}
where $m_{\text{conc}}(|{\bf r}|)$ is the profile in the concentric
geometry given by Eq.$\,$(\ref{Ib5}).

Let $D$ denote the surface-to-surface distance between the two spheres
and $R$ their common radius; then the corresponding radii $R_{\pm}$ in 
the concentric geometry are given by
\begin{equation}
\label{IRpm1}
R_{\pm}=\frac{D}{4}\,A\left(\frac{A+1}{A-1}\right)^{\pm 1}
\end{equation}
with the abbreviation
\begin{equation}
\label{IRpm3}
A=\sqrt{1+4\,\frac{R}{D}}\;.
\end{equation}
Putting the pieces together, Eqs.$\,$(\ref{Ib2})-(\ref{IRpm1}) 
yield the order parameter profile in the geometry of two spherical particles
with radius $R$ a distance $D$ apart from each other
right at the critical point $(t, h)=(0,0)$.

\subsection{Short distance expansion}
\label{sub_sde}
In the following we describe the methods for obtaining results off criticality,
i.e., for $t\gtrless 0$ and $h\neq 0$,
 i.e.,  $\tau\gtrless 0$ and $H\neq 0$.
Equation (\ref{Ib1}) will then yield a different profile 
$m({\bf r}^{\,\prime})$. However, the profile at the critical point
serves as a suitable starting point for the minimization procedure
(\ref{If3}).
Close to the surfaces it is even possible to calculate the
deviation of the order parameter
from the profile at the critical point 
analytically. In this context
the so-called short distance expansion turns out to be useful,
and has been carried out 
for $\tau\gtrless 0$
in ref. {}\cite{HD99}. With $H\neq 0$ in addition, it reads:
\begin{eqnarray}
\label{sde1}
m(s \to 0;R,\tau,H) & =&   \frac{\sqrt{2}}{s} -\frac{\sqrt{2}}{6}
\frac{d-1}{R}\\[0.5em]
&+&\frac{\sqrt{2}}{6}\,s\Big\{-\, \tau 
+\Big[ \frac{5}{6} \, \frac{(d-1)^2}{R^2} \, - \,
\frac{(d-1)(d-2)}{R^2} \Big]  \Big\}\nonumber \\[0.5em]
&+& \frac{\sqrt{2}}{4}\,s^2\Big(\frac{H}{\sqrt{2}}-\frac{d-1}{3\,R}\,\tau
-\frac{4d^3-39d^2+120d-85}{54\,R^3}\Big)\nonumber\\[0.5em]
&+&{\cal O}(s^3)\;,\nonumber
\end{eqnarray}
where $s$ is the radial distance from the surface of a sphere with
radius $R$.
The difference $\delta m=m(s \to 0;R,\tau,H)-m(s \to 0;R,\tau=0,H=0)$ 
gives the deviation from the profile at the critical point
close to the surfaces. While the profile itself diverges
at the surfaces, the deviation $\delta m$ vanishes
with the radial distance $s$ as
\begin{equation}
\label{sde2}
\delta m (s\to 0;R,\tau,H)=-\frac{\sqrt{2}}{6}\,\tau\,s
+\frac{1}{4}\left(H-\frac{\sqrt{2}}{3}\frac{d-1}{R}\,\tau\right)\,s^2
+ \ldots \;,
\end{equation}
where the bulk field $H$ and the radius $R$ of the spherical particles
appear only in ${\cal O}(s^2)$ and higher order.
Equation (\ref{sde1}) is only valid for small $s$ since the presence
of the second particle is neglected. Corrections to 
Eqs.$\,$(\ref{sde1}) and (\ref{sde2}), however, are of order
${\cal O}(s^3)$ as derived for the case of parallel plates,
where they are called distant wall corrections \cite{FdG78,FAY80}.

\subsection{Numerical solution for the complete profiles off criticality}
\label{sub_min}
\begin{figure}[h]
\setlength{\epsfxsize}{10cm}
\centerline{\epsfbox{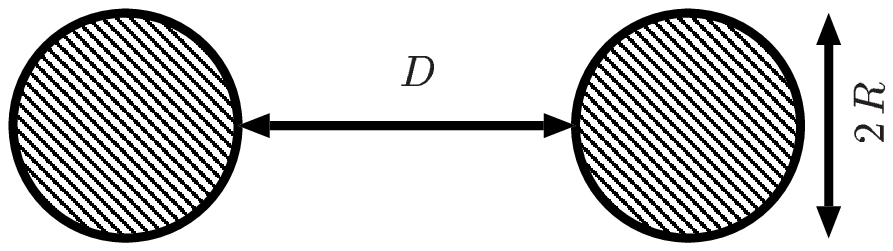}}
\caption{\label{fig_geo}The geometry of two equal neighboring spheres.}
\end{figure}

So far we have provided results for the order parameter 
profile at the critical point for the two sphere geometry
[Eq.$\,$(\ref{Ib9})] and off criticality
but close to the surface of a single sphere [Eq.$\,$(\ref{sde1})].
In order to obtain the complete order parameter profile away from the critical
point for the geometry consisting of two spheres with equal size 
(see Fig.$\:$\ref{fig_geo}) additional, numerical effort is required.

Upon discretization the order parameter profile 
$m({\bf r}^{\,\prime};D,R;\tau,H)$ 
in the two sphere geometry (Fig.$\,$\ref{fig_geo})
is described by a set of parameter values $a_{ij}$ where $i$ and $j$
indicate the dependence on the two independent spatial variables
required for the present cylindrical symmetry around the axis
between the two centers of the spheres [see, c.f., Eq.$\,$(\ref{param2})]:
\begin{equation}
\label{param1}
m({\bf r}^{\,\prime};D,R;\tau,H)=m(\{a_{ij}\})\;.
\end{equation}
Within mean-field theory ${\cal H}(\{a_{i,j}\})$ has to be
minimized with respect to $\{a_{ij}\}$ 
for given $D$, $R$, $\tau$, and $H$.
To this end we apply the method of steepest descent \cite{BF93},
\begin{equation}
\label{mosd1}
a_{ij}^{\,(n+1)}=a_{ij}^{\,(n)}-\kappa\,\frac{\partial\,{\cal H}(\{a_{ij}\})}
{\partial\,a_{ij}}
\Bigg|_{\textstyle a_{ij}^{\,(n)}}\;,
\end{equation}
where $\kappa$ is a positive number and $a_{ij}^{\,(n)}$ is the 
$n$-th iteration of the parameter $a_{ij}$.

It is known that in general the method of steepest descent converges
slowly. Nevertheless in the present case it is found to be useful 
since the dependence on the choice of the initial set of 
parameters $\{a_{ij}^{\,(0)}\}$ is weak.
In fact, it turns out that the profile at the critical point is
a rather useful starting profile. The convergence also depends on
the value of $\kappa$: if it is chosen too large, the convergence breaks
down, if it is chosen too small, the convergence becomes too slow.
As we will discuss later, one advantage of the present minimization
method is that the explicit evaluation of the Hamiltonian $\cal H$ is not
necessary, since only the gradient
$\frac{\partial\,{\cal H}(\{a_{ij}\})}{\partial\,a_{ij}}$
is needed, which can be obtained independently.
In order to keep this advantage, iterative methods to
determine the best choice for the value of $\kappa$, which
explicitly need the value of the functional that is to be
minimized, are not applied. Nonetheless the best choice of $\kappa$
is the largest value for which the minimization still proceeds.

In particular, we use the parametrization
\begin{equation}
\label{param2}
m({\bf r}^{\,\prime}_{ij};D,R;\tau,H)=m({\bf r}^{\,\prime}_{ij};D,R;0,0)
+a_{ij}\;,
\end{equation}
where ${\bf r}^{\,\prime}_{ij}$ denotes discretized spatial positions in the 
two sphere geometry, $m({\bf r}^{\,\prime}_{ij};D,R;0,0)$
is the value of the order parameter profile for $(\tau,H)=(0,0)$
at those positions, according to Eq.$\,$(\ref{Ib9}), and $a_{ij}$ 
denotes the deviation
from this value at the position ${\bf r}^{\,\prime}_{ij}$
used in the minimization
procedure (\ref{mosd1}). Thus starting the procedure with the profile 
for $(\tau,H)=(0,0)$ implies the choice
\begin{equation}
\label{param3}
a_{ij}^{(0)}=0\;,\quad\forall\;i,j\,.
\end{equation}
In addition, Eq.$\,$(\ref{sde2}) provides the behavior of the $a_{ij}$
for ${\bf r}^{\,\prime}_{ij}$ close to the surface, i.e., $a_{ij}^{(n)}=0$ 
for all $n$ if ${\bf r}^{\,\prime}_{ij}={\bf r}^{\,\prime}_S$. From
Eq.$\,$(\ref{sde2})
even the slope of a function $f(s)$ used for interpolating
the $a_{ij}$ as function of the radial distance $s$ from the surface of the
sphere is known for $s\to 0$.

The cylindrical symmetry of the problem allows one to consider
the planar geometry depicted in Fig.$\,$\ref{fig_geo}.
Instead of discretizing the space by a square lattice we use a lattice 
which takes into account that there are areas with strong changes in the
order parameter $m$, i.e., close to the surfaces, and areas with weak
changes, i.e., far away from the spheres. In those areas with weak
changes less lattice points are needed. For this purpose we use
what we call
the {\em conformal lattice}. The idea is to use the conformal mapping
(\ref{Ib7}) in order to create this lattice from 
that one for the case of concentric
spheres. In analogy to electrostatics
the field lines and the lines of equal potential
of a system with two concentric spheres forming the two electrodes
of a capacitor are mapped onto the two sphere geometry such that
the new lines are the field lines and the lines of equal potential
of a system with two charged spheres. The intersections of
these lines form the lattice sites of the conformal lattice.
The advantage of this lattice is that the field lines approach
the spheres {\em perpendicularly} to their surface, which allows one to 
use the short distance behavior [Eqs.$\,$(\ref{sde1}) and (\ref{sde2})]
as a boundary condition for the $a_{ij}$.
%
% Fig.3
%
Figure$\,$\ref{fig_cnfl} shows the conformal lattice for the
geometry of two spheres. In principle, 
it would be possible to use only one half of this lattice,
since this system is symmetric
with respect to the vertical midplane. However, this reduction leaves one
with a disadvantage. The value of the order parameter at this midplane
wall is not known, and thus it must be determined within the minimization
process. The minimization turns out to be inefficient then.
Therefore for the two sphere geometry we resort to the full conformal
lattice enclosing both spheres as shown in Fig.$\,$\ref{fig_cnfl}.

The Hamiltonian $\cal H$ [Eq.$\,$(\ref{I130})] is given by
\begin{equation}
\label{min1}
{\cal H}=\frac{6}{u}\int\limits_V dV {\cal L}(m)
+ \frac{6}{u}\int\limits_S dS {\cal L}_S(m)\;,
\end{equation}
with
\begin{equation}
\label{min2}
{\cal L}(m)=\frac{1}{2}(\nabla m)^2 +\frac{\tau}{2}m^2+\frac{1}{4}m^4-H\,m
\end{equation}
and ${\cal L}_S(m)=\frac{c}{2}\,m^2-H_1\,m$ 
with $H_1=h_1\,\sqrt{u/6}$.
Performing the functional derivative with respect to the order 
parameter $m$ leads, via partial integration, to
\begin{equation}
\label{min3}
\frac{\delta\,{\cal H}\{m({\bf r})\}}{\delta\,m({\bf r})}
=\frac{6}{u}\int_V\!dV\{-\Delta\,m+\tau\,m+m^3-H\} + \frac{6}{u}
\int_S\!dS\,\{-\partial_n\,m+{\cal L}^{\prime}_S(m)\}
\;,
\end{equation}
where $\partial_n\,m$ is the derivative of the order parameter
in direction of the surface normal. We note that the surface integral 
vanishes, if the
boundary conditions [Eqs. (\ref{sde1}) and (\ref{sde2})] are fulfilled.
Equation$\,$(\ref{mosd1}) contains the derivative of
the Hamiltonian ${\cal H}$ with respect to the parameters
$a_{ij}$ which represent the order parameter $m$. 
With $m=m(a_{ij})=const$ in a small volume  
$\delta V({\bf r}^{\,\prime}_{ij})$ around the
position ${\bf r}^{\,\prime}_{ij}$ the gradient in
Eq.$\,$(\ref{mosd1}) is approximated by
\begin{equation}
\label{min4}
\frac{\partial\,{\cal H}(\{a_{ij}\})}{\partial\,a_{ij}}=
\frac{6}{u}\,{\delta V({\bf r}^{\,\prime})}\,\{-\Delta\,m+\tau\,m+m^3-H\}
\Big |_{\displaystyle{\bf r}^{\,\prime}={\bf r}^{\,\prime}_{ij}}\quad.
\end{equation} 
Note that a constant prefactor in Eq.$\,$(\ref{min4}) can be
absorbed by a rescaling of $\kappa$ in Eq.$\,$(\ref{mosd1}).
If Eq.$\,$(\ref{Ib1}) holds, i.e.,
if the expression in curly brackets in Eq.$\,$(\ref{min4}) vanishes,
Eq.$\,$(\ref{mosd1}) reduces to
\begin{equation}
\label{mosd4}
a_{ij}^{\,(n+1)}=a_{ij}^{\,(n)}\;,
\end{equation}
i.e., the minimization has led to a fixed point solution
of the iteration.

For the minimization process
the second derivatives of the order parameter profile are needed
[see Eq.$\,$(\ref{min4})]. Since the order parameter is calculated only
at the discrete sites ${\bf r}^{\,\prime}_{ij}$, numerical
differentiation must be applied. For a numerical derivative,
in general, it is preferable to have the sites 
on an equidistant square lattice. However, it is also possible to
deal with derivatives on the conformal lattice. To this end we have used
the spline approximation \cite{S90} in order to interpolate 
between all the $a_{ij}$
on each of the lines of the conformal lattice (see Fig.$\,$\ref{fig_cnfl}).
This approximation directly yields the second derivative of the
order parameter at the lattice sites with respect to
the variables parametrizing the lines. Under a conformal mapping
angles are conserved, so that the lines in the conformal lattice
intersect perpendicularly which allows one 
to translate the derivatives obtained from the spline approximation
into derivatives with respect to the cylindrical coordinates 
${\bf r}^{\,\prime}=(\rho,z)$.

\subsection{Stress tensor}
\label{sub_st}
With the tools now available to determine the order parameter profile in
the two sphere geometry
the next step is to calculate the effective forces between the two spheres.
The straightforward way to determine such forces is to calculate the 
free energy
$\displaystyle -k_BT\,\ln{\cal Z}$ as function of the distance $D$
between the spheres,
and then to estimate the derivative with respect to the distance
by considering finite differences. However, the total
value of the free energy is expected to be large as compared with the
differences,
and thus one has to deal numerically with differences between
large numbers, which causes high inaccuracies. Therefore we resort to
an alternative method, in which the force is calculated {\em directly} 
using the so-called stress tensor ${T}_{\mu\nu}$, which is
defined by the linear response
\begin{equation}
\label{ST1}
\delta{\cal H} = \int_{V} d^dr \: \frac{\partial\,
b_\mu}{\partial x_\nu}  \: {T}_{\mu\nu}({\bf r})\;,
\end{equation}
where ${\bf b}$ is a non-conformal coordinate transformation,
$\mu$ and $\nu$ index the spatial coordinates,
and $\delta{\cal H}$ is the corresponding energy shift.
With the Lagrangian ${\cal L}(m)$ in Eq.$\,$(\ref{min2})
it reads \cite{BE95,HSED98,B80}
\begin{equation}
\label{ST2}
{T}_{\mu\nu}=\frac{6}{u}\left[\frac{\partial{\cal L}}
{\partial(\partial_\nu m )}
\partial_\mu m -\delta_{\mu\nu}{\cal L}\right]\;.
\end{equation}
The knowledge of the stress tensor on {\em any} infinitely extended
surface which separates the two spheres from one another, or likewise
on any closed, finite surface which contains only one of the spheres,
allows one to calculate the force between the spheres.
This is illustrated in terms of the spherical volume $V_S$ with
radius $\widetilde{R}$ containing
the right sphere in Fig.$\,$\ref{fig_sts}. Using the coordinate transformation
\begin{eqnarray}
b_z & = & \alpha\,,\;\forall\,{\bf r}\in V_S\\
&=& 0\,,\;\forall\,{\bf r}\not\in V_S\nonumber\\
b_{\nu\neq z}&=&0\;,\nonumber
\end{eqnarray}
the volume $V_S$ is shifted in $z$-direction by the amount
$\alpha$. The derivative $\frac{\partial\,
b_\mu}{\partial x_\nu}$ then leads to a delta function such
that the integral over the whole volume $V$ reduces to
an integral over the boundary of $V_S$. 
In this case the force $F_{\text{sing}}=\frac{k_B\,T_c}{R}\,K$
with the universal scaling function $K$ 
[see Eq.$\,$(\ref{scalf})] 
is given by
\begin{eqnarray}
-\frac{\displaystyle\partial\,\delta{\cal H}}{\partial\alpha}
& = & -\frac{\displaystyle\partial}{\partial\alpha}\int_{V} d^dr
\: \frac{\displaystyle\partial b_\mu}{\partial x_\nu} \: T_{\mu\nu} \nonumber\\
& = & -\int_{\partial V_S}   d\Omega \; n_\nu \: T_{z\nu}
\label{KglInt}\\
& = & -4\pi\widetilde{R}^3 \int^\pi_0 d\varphi \sin^2\varphi\,
 [T_{zz}(\widetilde{R},\varphi)\,\cos\varphi + T_{z\rho}(\widetilde{R},\varphi)
\,\sin\varphi]\quad,
\nonumber
\end{eqnarray}
where ${\bf n}$ is the unit vector perpendicular to the boundary
$\partial V_S$.
The quantities used in the parametrization in Eq.$\,$(\ref{KglInt})
are sketched in Fig.$\,$\ref{fig_sts}.

%
% Fig.4
%

In the geometry with  cylindrical symmetry [$m=m(\rho,z)$]
considered here, Eq.$\,$(\ref{ST2}) leads to
\begin{equation}
\label{Tzz}
T_{zz} = \frac{1}{u} \left [3 \left (\frac{\partial m}{\partial z}
\right )^2 +
4 \frac{m}{\rho} \frac{\partial m}{\partial \rho} +
2 m \frac{{\partial}^2 m}{{\partial \rho}^2} - \left
(\frac{\partial m}{\partial \rho} \right )^2 - 3\tau m^2 -\frac{3}{2}
m^4 \right ]
\end{equation}
and
\begin{equation}
\label{Tzrho}
T_{z\rho} = \frac{1}{u} \left [4 \left (\frac{\partial m}{\partial
z} \right ) \left (\frac{\partial m}{\partial \rho} \right ) 
- 2 m \frac{{\partial}^2 m}{ {\partial \rho} {\partial z} } \right ]\;,
\end{equation}
where the so-called improvement term $-\frac{1}{u}(\nabla_{\mu}\nabla_{\nu}
m^2-\delta_{\mu\nu}\Delta m^2)$
has been added \cite{B80}.

In our calculation, we choose $V_S$ as indicated in Fig.$\,$\ref{fig_cnfl},
determine the stress tensor at the lattice points, and use
spline interpolations between them. Note that the possibility of
choosing different spherical surfaces enclosing one of the spheres,
which via Eq.$\,$(\ref{KglInt}) should all result in the same force 
$F_{\text{sing}}$,
provides a valuable and sensitive check for both the validity and the 
accuracy of the numerical results.

%%%%%%%%%%%%%%%%%%%%%%%%%%%%%%%%%%%%%%%%%%%%%%%%%%%%%%%%%%%%%%%%%%
\section{Results for the two-particle interaction}
\label{sec_ses}
The goal of this work is to provide a major step towards the understanding
of aggregation of colloidal particles dissolved in a binary liquid
mixture close to criticality.
In the previous section we explained the various methods for obtaining
the critical Casimir force between two spherical particles of equal size. 
The results are presented in this section.

The aggregation phenomenon is driven by the dependences on 
the temperature $t=(T-T_c)/T_c$
and on the bulk composition $c_A$ of the solvent.
While the former is related
to the temperature parameter $\tau$ the latter is related to the bulk 
field $H$. Thus also the Casimir force between two particles
must be determined as a function of those two parameters.
We present results for the scaling function 
$K_{\pm}(\Delta=\frac{D}{R},\Theta_{\pm}=\frac{D}{\xi_{\pm}},
\Xi=\text{sgn}(h)\,\frac{D}{l_H})$
which is related to the force $F^{\pm}_{\rm sing}$ between the particles
via Eq.$\,$(\ref{scalf}).
The correlation length $\xi_{\pm}$ for $H=0$ depends on the temperature $T$ 
as [see also Eq.$\,$(\ref{If1})]
\begin{equation} 
\xi_{\pm}=\xi^{\pm}_0\,|t|^{-\nu}=\left\{ 
\begin{array}{r@{\quad}l}
\displaystyle{\tau^{\,- 1/2} \, \, \, \, ,} & {\tau > 0 } \\
\displaystyle{|2\,\tau|^{\,- 1/2} \, \, \, ,} & \tau < 0 \, \,
\end{array}
\right.\;,\quad H=0\;,
\label{xipm}
\end{equation}
with $\nu=\frac{1}{2}$ within mean-field theory
while the correlation length $l_H$ at $T_c$ depends on the bulk field $H$
as
\begin{equation} 
\label{lH}
l_H=l_0\,\left|\frac{c_A-c_{A,c}}{c_{A,c}}\right|^{-\nu/\beta}
=\frac{1}{\sqrt{3}}\,{|H|}^{\,-{1}/{3}}\;,\quad\tau=0\;,
\end{equation}
where $\nu/\beta=1$ within mean-field theory;
$\xi^{\pm}_0$ and $l_0$ are nonuniversal amplitudes.
The first parts of Eqs.$\,$(\ref{xipm}) and (\ref{lH}) hold in
general whereas the second parts correspond to the present mean-field
description.
The scaling variable $\Xi=\text{sgn}(H)D/l_H$ depends on the 
sign of $H$ [Eq.$\,$(\ref{scalf})].
The relation between the
bulk field $H$ and the concentration $c_A$ will be discussed further
in Subsec.$\,$\ref{sub_eos},
but we keep in mind that a negative value of $H$ means that the fluid 
component, which is preferred by the particles, is the one with a
low concentration in the bulk (see Fig.$\,$\ref{fig_pdp}).   

\subsection{Dependence on temperature at the critical composition}
\label{sub_tsn}
%
% Fig.5
%
The variation of the temperature leads to an interesting behavior of
the force \cite{HSED98}. 
According to Fig.$\,$\ref{fig_tscn} the force exhibits 
a {\em maximum} for a temperature above the critical point. 
The position of the maximum of the force as 
function of the scaling variable $\Theta_+$ depends on the distance $D$ between
the particles. In the limit of large distances the position 
$\Theta_{\rm max}=\Theta_{\rm max}(\Delta,\Xi=0)$ of the maximum 
approaches that one of the critical point,
i.e.,
\begin{equation}
\Theta_{\rm max}\left(\Delta=\frac{D}{R} \to\infty\right)\to 0\;.
\end{equation}
On the other hand, for small distances the position of the maximum approaches 
a finite value {\em above} $T_c$
\begin{equation}
\Theta_{\rm max}\left(\Delta\to 0\right)\to\Theta_{\rm max}^{(0)}\;.
\end{equation} 
These two limiting behaviors are shown in
Fig.$\,$\ref{fig_tscn}. The disappearance of the maximum for large 
distances along with the derivation of the {\em small radius expansion}
will be discussed in more detail later on in this section.

\subsection{Dependence on composition at the critical temperature}
\label{sub_hsn}

%
% Fig.6
%

\begin{figure}[h]
\setlength{\epsfxsize}{6cm}
\centerline{\epsfbox{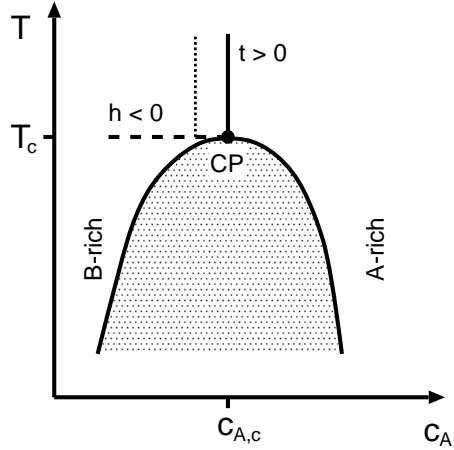}}
\caption{\label{fig_pdp}Schematic phase diagram of a binary liquid mixture
consisting of $A$ and $B$ particles in terms of temperature $T$
and concentration $c_A$ of $A$ particles.
The shaded area is the two phase region $-$ ending in the critical point $CP$
$-$ which separates the A-rich
and the B-rich phase. The relevant thermodynamic variables 
near the critical point are $t$ and $h$.
Three different thermodynamic paths are indicated as considered
in the main text.}
\end{figure}

The above results show that there are rich structures in the force curves off
criticality. Therefore it is important to monitor the force
not only at the critical point, but also in its vicinity.
This becomes even more evident
if not the temperature but the composition of the solvent is varied.
Along the dashed path in Fig.$\,$\ref{fig_pdp} and for small 
distances $D$ the force exhibits a pronounced maximum as can be seen 
in Fig.$\,$\ref{fig_hscn}. 
For large distances there is only a maximum at the
critical point, i.e., $H=0$. 
For $\Xi>0$ the force decays monotonously for all distances $D$. 

%
% Fig.7
%

Figure$\,$\ref{fig_maxf} shows the maximum values of the force
for a given distance $D$ as function of
the temperature (solid line in Fig.$\,$\ref{fig_pdp}) or the 
composition (dashed line in Fig.$\,$\ref{fig_pdp}).
For comparison, also the force at the critical
point for the same particle distance $D$ is plotted 
(dash-dotted line in Fig.$\,$\ref{fig_maxf}). 
For intermediate and larger distances the three lines merge, which
is due to the fact that for large distances the position of the
maximum of the force approaches that of
the critical point. For small distances the differences are pronounced.
Obviously, the composition of the solvent plays an important role.

\subsection{Dependence on temperature off the critical composition}
\label{sub_grn}

%
% Fig.8
%

So far the thermodynamic paths were chosen such that either the 
temperature or the solvent 
composition were fixed at their critical point value. However, 
it is experimentally relevant to consider also thermodynamic paths 
along which the temperature is varied at fixed compositions
$c_A\neq c_{A,c}$. The corresponding results are given 
in Fig.$\,$\ref{fig_escn} for a fixed distance $D$. 
For small deviations $|c_A - c_{A,c}|\ll  c_{A,c}$ the position of 
the maximum of 
the Casimir force is more or less unchanged while its absolute value
is changed considerably.
However, the main features of the temperature scan
do not differ much from that at $c_{A,c}$ so that they are sufficiently robust
in view of possible experimental tests with finite resolution for $c_A$.

If the dotted path in Fig.$\,$\ref{fig_pdp} is
extended to temperatures below $T_c$ the force is analytic as function of
the temperature until the path intersects with the coexistence line.

\subsection{Derjaguin approximation}
\label{sub_der}
In the limit that the radius $R$ of the spherical particles
is much larger than both the correlation length $\xi$ and the distance $D$ 
of closest approach surface-to-surface between the two  particles, 
the particles can be regarded as composed of
a pile of fringes. Each fringe builds a fringe-like slit with distance
$L = D + r_{||}^2 / R $, where $r_{||}$ is the radius of the
fringe. Within this Derjaguin approximation \cite{D34} and in $d = 4$,
the force between the particles 
in units of $\frac{k_BT_c}{R}$ [see Eq.$\,$(\ref{scalf})]
is given by
\begin{equation}
\label{derj}
{K_{\pm}\left(\Delta,\Theta_{\pm},\Xi\right)} =  4\,\pi\,
\Delta^{-{5}/{2}}\int_0^{\infty} \!\!\!\!\! d\upsilon\,\upsilon^2
\;\frac{\displaystyle 
K_{\pm}^{||}\!\left(\Theta_{\pm}\,(1+{\upsilon^2}),
\Xi\,(1+{\upsilon^2})\right)}
{(1+{\upsilon^2})^{4}}\,, 
\end{equation}
where $\upsilon$ is a dimensionless integration variable and 
$K_{\pm}^{||}(L/\xi_{\pm},\text{sgn}(H)L/l_H)$ is the analogous 
scaling function for the Casimir force in
the slit geometry with parallel walls at distance $L$
[see, c.f., Sec.$\,$\ref{sec_flm}].

\subsection{Small radius expansion}
\label{sub_sre}

If on the other hand the 
radius $R$ $-$ albeit large on the microscopic scale
$-$ is much {\em smaller\/} than 
$D$ and $\xi$, the statistical Boltzmann weight 
$e^{- \delta {\cal H}_S}$ characterizing the presence of the
sphere centerd at ${\bf r}_S$ can be systematically expanded 
in terms of increasing powers of $R$ \cite{BE95}, i.e.,
\begin{equation} \label{s10}  
e^{- \delta {\cal H}_S}/d_{\uparrow}=
1 + c_{\,\uparrow}^{\Phi} \, R^{\,x_{\Phi}} \, \Phi({\bf r}_S)
+ c_{\,\uparrow}^{\Phi^2} R^{\,x_{\Phi^2}} \, \Phi^2({\bf r}_S) + \ldots
\end{equation}
where $d_{\uparrow}$ is an amplitude and where
$x_{\Phi} = \beta / \nu$ and $x_{\Phi^2} = d - \nu^{-1}$ 
(with the standard bulk critical exponents $\beta$ and $\nu$) are the scaling 
dimensions of $\Psi = \Phi, \Phi^2$. The ellipses stand for 
contributions which vanish more rapidly for $R \to 0$.
The coefficients $c_{\,\uparrow}^{\Psi}$ are {\em fixed\/} by
$c_{\,\uparrow}^{\Psi} = A_{\,\uparrow}^{\Psi} / B_{\Psi\,}$, where
$A_{\,\uparrow}^{\Psi}$ and $B_{\Psi}$, respectively, are
amplitudes of the half-space (hs) profile 
$\langle \Psi(z) \rangle_{{\rm hs}, \, T = T_c}^{\,\uparrow} =  
A_{\,\uparrow}^{\Psi\,} (2 z)^{-x_{\Psi}}$ 
{\em at\/} the critical point of the fluid for the boundary 
condition $\uparrow$ corresponding to critical adsorption,
and of the bulk two-point correlation function
$\langle \Psi({\bf r}) \Psi(0) \rangle_{b, \, T = T_c} = 
B_{\Psi\,} r^{-2 x_{\Psi}}$ \cite{BE95}. 
Since both spheres are equally small the free energy $\cal F$ follows
from the bulk average of two statistical weights of the type given in
Eq.$\,$(\ref{s10}):
\begin{equation} \label{sre2}  
\frac{{\cal F}(D) - {\cal F}(\infty)}{k_B T_c}
= - \ln \left[
\frac{ \langle e^{- \delta {\cal H}_{S_1}}
e^{- \delta {\cal H}_{S_2}}\rangle_{\rm b} }
{\langle e^{- \delta {\cal H}_{S_1}} \rangle_{\rm b}
\langle e^{- \delta {\cal H}_{S_2} }\rangle_{\rm b} } \right] \, \, ,
\end{equation}
where the spheres $S_1$ and $S_2$ are a distance $D$ apart from 
each other. Using Eq.$\,$(\ref{s10}) we find \cite{HSED98}
\begin{equation} 
\label{sre4}
\frac{{\cal F}(D) - {\cal F}(\infty)}{k_B T_c}
=-(c_{\,\uparrow}^{\Phi} \, R^{\,x_{\Phi}})^2
\langle\Phi({\bf r}_{S_1})\,\Phi({\bf r}_{S_2})\rangle_{\rm b}
\, [1 - 2 c_{\,\uparrow}^{\Phi} \, 
R^{\,x_{\Phi}} \langle \Phi \rangle_{\rm b} + \ldots ]\, ,
\end{equation}
where $\langle\Phi({\bf r}_{S_1})\,\Phi({\bf r}_{S_2})\rangle_{\rm b}$
is the bulk two point correlation function
and $\langle \Phi \rangle_{\rm b}$ provides a measure of the deviation 
of the bulk composition from its value at the critical point.
The Casimir force is the derivative of the free energy with respect
to the distance $D$ between the spheres. Note that in Eq.$\,$(\ref{sre4})
the first term in the
square brackets gives rise to an attractive force which is 
{\em increased\/} by the second term if $\langle \Phi \rangle_{\rm b}$
is negative, i.e., if the binary liquid mixture is poor in the 
component preferred by the colloids 
(compare Subsecs.\,\ref{sub_hsn} and \ref{sub_grn}).

Within mean-field approximation, the correlation function is given by
(see Eq.$\,$(B5) in ref. {}\cite{EHD96})
\begin{eqnarray}
{\langle\Phi({\bf r}_{S_1})\,\Phi({\bf r}_{S_2})\rangle}_{b}
&=&G_b^{+}(|{\bf r}_{1}-{\bf r}_{2}|=D;\xi_+)\nonumber\\ 
&=&B_{\Phi}\frac{1}{D^{d-2}}\,
\underbrace{
\frac{2^{2-d/2}}{\Gamma(\frac{d-2}{2})}\,
\left(\frac{D}{\xi_+}\right)^{\frac{d-2}{2}}\,
\text{K}_{\frac{d-2}{2}}(D/\xi_+)}\;,\;
\label{sre7}\\
&&\qquad\qquad\qquad\qquad \to 1\:,\;T\to T_c\nonumber
\end{eqnarray}
where $\text{K}_{(d-2)/2}$ is a modified Bessel function and
where $\xi_+= l_H$ for the case $\tau=0$ and $H\neq 0$.
Thus as expected two point-like perturbations
of the bulk do not exhibit a rich interaction structure
but a monotonous decay.
If a small sphere is immersed at a distance $D$ from another small
sphere with a given adsorption profile, the resulting perturbation
of this profile leads to an effective interaction.
At those distances, at which the small radius expansion can be applied
the profile of the distant sphere is already close to the bulk value.
This differs from the system studied in ref. {}\cite{HSED98} where
a small sphere in front of a planar wall is considered. In that case
the sphere is exposed to the half-space adsorption profile 
which is a much stronger perturbation of the bulk order
parameter than that of a sphere (compare refs. {}\cite{HD99} and
{}\cite{Han2000}).

\subsection{Comparison with dispersion forces}
\label{sub_dis}

It has been stressed above that the Casimir forces depend sensitively
on both temperature and composition of the solvent. The Casimir forces
add to the omnipresent dispersion forces, which also depend on the 
thermodynamic state of the solvent. However, this latter dependence 
is smooth and - within the critical region of interest here - it is weak.
Therefore the dependence of the Casimir force on temperature and concentration
allows one to extract it from the actual total force which is the only one
which is experimentally accessible.
As an example for a binary liquid mixture we consider a water-lutidine
mixture for which the critical temperature is $T_c=307.1\;\text{K}$ \cite{BE}.
The strength of the dispersion forces
is characterized by the Hamaker constant, 
which for a system of silica spheres immersed 
in such a mixture is approximately $A=10^{-20}\;\text{J}$.
Figure$\,$\ref{fig_frl} shows the comparison of the dispersion
forces and of the Casimir forces at the critical point for spheres
with a typical radius $R=100\;\text{nm}$ which add up
to the total force 
\begin{equation}
\label{ftot}
F_{\text{tot}}(\Delta, 0 ,0)=\frac{k_BT_c}{R}\,K_+(\Delta, 0, 0)
+F_{\text{disp}}(\Delta)\;.
\end{equation}
The dispersion forces are given by \cite{CA96}
\begin{equation}
\label{fdisp}
F_{\text{disp}}(\Delta)=-\frac{A}{R}\frac{32}{3}\frac{1}{(\Delta+2)\,(\Delta^2+4\,\Delta+4)\,(\Delta+4)^2\,\Delta^2}\;,
\end{equation}
where $A$ is the aforementioned Hamaker constant. 
In the limit of small distances the Casimir force is known from the 
Derjaguin approximation
\begin{equation}
\label{Ksmall}
K_+(\Delta\to 0,0,0)=-\pi\,\Delta_{\uparrow\uparrow}\,\Delta^{-2}\;,
\end{equation}
where $\Delta_{\uparrow\uparrow}=0.326$ is the estimate
of the Casimir amplitude in $d=3$ \cite{K97},
and in the limit of large distances the Casimir force can be obtained from the
small radius expansion \cite{ER95}
\begin{equation}
\label{Klarge}
K_+(\Delta\to\infty,0,0)=-2\,\frac{\beta}{\nu}\,{\cal A}_{\uparrow\uparrow}
\,(\Delta+2)^{-2\,{\beta/\nu}\,-1}\;,
\end{equation}
where ${\beta/\nu}=0.52$ and 
${\cal A}_{\uparrow\uparrow}=\frac{(A^{\Phi}_{\uparrow})^2}{B_{\Phi}}=7.73$
\cite{HSED98}.

%
% Fig.9
%

In Fig.$\,$\ref{fig_frl} also the forces between a single sphere
and a planar wall are shown, as considered in ref. {}\cite{HSED98}.
The dispersion forces then read 
$F_{\text{disp}}(\Delta)=-\frac{A}{R}\frac{2}{3}
\frac{1}{(\Delta+2)^2\,\Delta^2}$, where $\Delta$ is the surface-to-surface
distance in units of the radius $R$. For small distances the amplitude
of the Casimir force is two times bigger than in the case of two spheres
while for large distances the small radius expansion leads
to \cite{ER95}
\begin{equation}
\label{KlarW}
K_+(\Delta\to\infty,0,0)=-2\,\frac{\beta}{\nu}\,{\cal A}_{\uparrow\uparrow}
\,(2\,\Delta)^{-{\beta/\nu}\,-1}\;.
\end{equation}
For small distances all forces increase as $\Delta^{-2}$, only the amplitudes 
differ from each other. The values considered for $A$ and $T_c$
lead to Casimir forces that are about ten times larger than the
dispersion forces.
For large distances, however, the (unretarded) dispersion forces 
$(\sim\Delta^{-7})$
decay faster than the Casimir forces $(\sim\Delta^{-2.04})$.
(Ultimately retardation of the dispersion forces leads to a decay 
$\sim\Delta^{-8}$ and $\sim\Delta^{-5}$ for SS and SW, respectively.
For $\Delta=1$ the Casimir forces between the two spheres are about
fifty times larger than the dispersion forces such that the latter
contribution to the total force [Eq.$\,$(\ref{ftot})] can be neglected.
For a sphere in front of a planar wall and for large distances
the Casimir forces 
$(\sim\Delta^{-1.52})$ decay even slower than they increase 
for small distances $(\sim\Delta^{-2})$.

\subsection{Equation of state}
\label{sub_eos}
For the demixing transition of a binary liquid mixture the 
difference $(\mu_A-\mu_B)-(\mu_A-\mu_B)_c$
of the chemical potentials of the two components is proportional to the bulk
field $h$. The potential difference leads to a concentration 
$c_A\neq c_{A,c}$ of species $A$ in the bulk.
The bulk order parameter $M_b$ is proportional to the concentration 
difference $M_b={\cal A}(c_A-c_{A,c})$, where the amplitude $\cal A$
depends on the actual definition of the order parameter.
For such a system in the vicinity of its critical
point the equation of state is given by \cite{W65}
\begin{equation}
\label{eos}
h= {\cal D} M_b^{\delta}
\,f^{}\left(t\,\left|\frac{B}{M_b}\right|^{1/\beta}\right)\;,
\end{equation}
where $\beta$ and $\delta$ are the standard critical exponents,
${\cal D}$ and $B$ are nonuniversal amplitudes, and the scaling function 
$f^{}(x)=1+ x$ describes the crossover between the critical behavior
at $t=0$ and $h=0$, respectively.
Eq. (\ref{eos}) yields $M_b=\pm B|t|^{\beta}$ for $h\to 0$ and 
$M_b=(\frac{h}{\cal D})^{1/\delta}$ for $t\to 0$ as limiting cases.
The choice of the amplitude $\cal D$ defines a field $h$. Since,
however, the combination $e^{\int dV h\,M_b}$ is the field
contribution of the statistical weight of the system, $\cal D$ is
fixed by the amplitude of the chosen order parameter.

Away from the critical point the two point correlation function decays
for large distances $r$ exponentially as 
$G_b(r)\sim{\exp(-r/\xi)}/{r^{(d-2)/2}}$,
with the true correlation length $\xi$.
Within mean-field theory,
where one has $\beta=\frac{1}{2}$ and $\delta=3$,
this correlation length is equivalent to
that one derived from the second moment of the correlation function,
which in momentum space is given by
$\tilde{G}_b(q)\sim\frac{1}{q^2+\xi^{-2}}$.
From this relation one infers \cite{D86}
\begin{equation}
\label{ximf}
\xi=\left(\tau+3\,m_b^2\right)^{-1/2}\;,
\end{equation}
with $m_b$ obtained from the mean-field equation of state:
\begin{equation}
\label{eosmf}
H=\tau\,m_b+m_b^3\;.
\end{equation}
Equations (\ref{ximf}) and (\ref{eosmf}) lead to the mean-field expressions
of the correlation lengths $\xi_{\pm}$ and $l_H$ in Eqs.$\,$(\ref{xipm}) 
and (\ref{lH}), respectively.
The correlation lengths are experimentally accessible by
measuring the scattering structure factor.
This allows one to determine the nonuniversal amplitudes $\xi_0^{\pm}$ and
$l_0$ in  Eqs.$\,$(\ref{xipm}) and (\ref{lH}).
By measuring the coexistence curve $M_b=\pm B|t|^{\beta}$ 
the nonuniversal amplitude $B$ is obtained.

However, the amplitude $l_0$ in Eq.$\,$(\ref{lH}) is
fixed once $\xi_0^+$ and the nonuniversal 
amplitude $B$ are known:
\begin{equation}
\label{l0}
l_0=\xi_0^+\,\left(\frac{B}{{\cal A}\,c_{A,c}}\right)^{\nu/\beta}\,
\left(\frac{Q_2}{\delta\,R_{\chi}}\right)^{\nu/\gamma}\;,
\end{equation}
where the first term in parantheses is nonuniversal
- but it does not depend on the definition of the order parameter -
and $Q_2$ and $R_{\chi}$ are universal amplitude ratios \cite{TF73} 
leading to
$\left({Q_2}/{\delta\,R_{\chi}}\right)^{\nu/\gamma}\approx 0.38$
in $d=3$.

Using Eq.$\,$(\ref{xipm})
and $m_b=\sqrt{|\tau|}$ one finds a relation
between the order parameter $m_b$ in the present model
and the order parameter $M_b$ defined in an actual experiment:
\begin{equation}
\label{mbMb}
m_b=\frac{1}{(\sqrt{2}\,\xi_0^-)^{\beta/\nu}}\,\frac{M_b}{B}\;.
\end{equation}
We note that in a magnetic system the bulk field $H$ is directly
accessible, while for liquids we use the correlation length $l_H$
as a measure of the bulk field $H$ according to Eq.$\,$(\ref{lH}).
So far we have assumed that the coexistence curve close to the critical
demixing point is symmetric with respect to the temperature axis.
However, this might not be the case in a real liquid mixture,
so that one would then have to use appropriate linear combinations
of the thermodynamic variables.

%%%%%%%%%%%%%%%%%%%%%%%%%%%%%%%%%%%%%%%%%%%%%%%%%%%%%%%%%%%%%%%%%%
\section{Results for the film geometry}
\label{sec_flm}

%
% Fig.10
%

%
% Fig.11
%
%
% Fig.12
%

%
% Fig.13
%
%
% Fig.14
%

In the previous section [see Eq.$\,$(\ref{derj}) in Subsec. \ref{sub_der}],
within the Derjaguin approximation, we have used results obtained for the 
film geometry in order to analyze the limiting case of
the colloidal particles being very close to each other.
Besides that the film geometry is of considerable interest in its own
right as the generic case for simulations \cite{WK98,FLB98} and for 
experimental
studies involving wetting films \cite{LPB98,GC} or force microscopes
with crossed cylinders of large radii of curvature. The Casimir forces 
in the film geometry have been considered
for walls that do not break the symmetry of the order parameter \cite{K91}
and also for symmetry breaking walls \cite{K97}. However, most studies
focus on the temperature dependence only, while in the following
we consider also the case $H\neq 0$.

For the film geometry the same numerical methods can be used as described in
Sec.$\,$\ref{sec_num}. 
However, since the system is
effectively one-dimensional there are additional methods available.
For instance, the first integral of the second-order differential equation 
(\ref{Ib1}) can be found analytically \cite{K97}. The integration constant, 
which follows from fulfilling the boundary conditions is proportional to 
the force (see App. A in ref. {}\cite{K97}). 
For $H=0$ the second integration
leads to elliptic integrals, while for $H\neq 0$ it must be carried out
numerically.

First we obtain the scaling function 
$K_{\pm}^{||}\!\left(\frac{L}{{\xi}},\,\frac{L}{{l_H}}\right)$
of the singular contribution to the force between two parallel plates at 
distance $L$ for the various thermodynamic paths shown
in Fig.$\,$\ref{fig_pdp}.
The temperature dependence (solid path) of the force is discussed
in refs. {}\cite{K97} and {}\cite{HSED98} 
(see in particular Fig.$\,$3 in ref. {}\cite{HSED98})
and reveals a maximum of the force at $\Theta_{\text{max}}=L/\xi_+=1.94$. 
As a new result the dependence of the force on the bulk field 
$H$ (dashed path) is shown in 
Fig.$\,$\ref{fig_kvh} in terms of the scaling variable
$\Xi=\text{sgn}(H)\,L/l_H$.

It is remarkable that the force maximum is about ten times larger than
the force at the critical point.
Since this scaling function enters into the expression for the force between
two spheres this peak structure carries over to that case, too.
Indeed, Fig.$\,$\ref{fig_maxf}
shows that the force maximum in the limit of two very close spheres 
is about five times higher than the force at the critical point.

Figure$\,$\ref{fig_cc} shows the dependence of the force on the bulk 
field for $T\neq T_c$. Above $T_c$ one finds a
similar structure like at $T_c$ except that the maximum is less pronounced.
(Here we refer to the maximum value of the modulus of the force. However,
in order to facilitate an easy comparison with the results of
ref. {}\cite{DME00}
in Fig.$\,$\ref{fig_cc} the sign of the force is chosen such that it 
reveals its attractive nature.)
Slightly below $T_c$ the maximum is more pronounced although the form
of the curve is still the same.
But at still lower temperatures the character of the curve changes. Instead 
of a smooth increase upon increasing the bulk field towards $H=0$
there is a jump from a small value of the force 
to a high one. This is accompanied by a first-order transition of
the corresponding order parameter 
profile.

Figure$\,$\ref{fig_trns} shows
the two coexisting profiles at the critical field of this first-order 
phase transition for various temperatures $T<T_c$. 
The profiles, which are negative in the middle of the  film,
correspond to the weaker of the two respective forces. This means that the 
forces are weaker if the solvent component preferred by the surfaces
is the one in which the bulk solution is poor.
These first-order phase transitions correspond to capillary condensation 
in the slit which can occur for fixed temperature as function of $H$
as shown in Figs.$\,$\ref{fig_cc} and \ref{fig_trns} or as function of 
temperature for fixed values of $H$. The loci of these capillary
condensation phase transitions are shown in Fig.$\,$\ref{fig_ccpd}.
The transition line ends in the capillary critical point 
($\Theta_-=7.17,\Xi=-9.54$)
which is in quantitative accordance with the findings for the 
critical point shift \cite{NF83}.

In ref. {}\cite{DME00} the solvation forces between two parallel
plates confining a two-dimensional Ising spin system have been calculated.
They exhibit the same features as those found in the present 
mean-field calculations valid for spatial dimensions $d\geq 4$.
For temperatures far below $T_c$
upon increasing the field the force jumps from a small to a large absolute
value and the varies linearly as function of the bulk field
(compare Fig.$\,$\ref{fig_cc}). 
In the linear regime the solvation force is given approximately by
${K_{-}^{||}}/{L^4}=2\,\bar{m}(T)\,H$ \cite{DME00},
where the bulk spontaneous
magnetization $\bar{m}(T)$, i.e., the magnetization at $H=0$ 
is $\bar{m}=|\tau|^{1/2}$ within 
mean-field approximation which indeed gives the slope of the linear
parts in Fig.$\,$\ref{fig_cc}. 
We note that in $d=2$ capillary condensation can be identified only as
a quasi-phase-transition whereas in $d=4$ there is indeed a bona fide 
first-order phase transition.
Figure$\,$\ref{fig_df} shows
the difference $\delta K_{-}^{||}$ between the two corresponding force values 
at the transition.
Upon approaching the capillary critical point this difference vanishes.
Along the coexistence line $H(T)$ of capillary condensation the quantity
$2\,\bar{m}(T)\,H(T)$ provides a good account of the force difference
(see Fig.$\,$\ref{fig_df}).

The findings in $d=2$ and $d=4$ agree to a large extent qualitatively.
Quantitatively, however, the results in $d=2$ and $d=4$ differ significantly.
For example the ratio of the maximum force and of the force at the critical
point as function of the bulk field and for $T=T_c$ is 
${K_{+}^{||}(0,\Xi_{\text{max}})}/{K_{+}^{||}(0,0)}\approx 100$ in
$d=2$ and ${K_{+}^{||}(0,\Xi_{\text{max}})}/{K_{+}^{||}(0,0)}\approx 10.1$ 
in $d=4$. A simple linear interpolation yields an estimated
value of ${K_{+}^{||}(0,\Xi_{\text{max}})}/{K_{+}^{||}(0,0)}\approx 55$ 
for this ratio in $d=3$. Thus the studies in 
ref. {}\cite{DME00} and the present ones can be used to obtain quantitative
estimates for the actual behavior in $d=3$.

%%%%%%%%%%%%%%%%%%%%%%%%%%%%%%%%%%%%%%%%%%%%%%%%%%%%%%%%%%%%%%%%%%
\section{Concluding remarks and summary}
\label{sec_con}

We have studied the critical adsorption profiles and the 
effective free energy of interaction for 
a pair of colloidal particles which are immersed in a binary
liquid mixture near its critical demixing point. Our results
pertain to the whole vicinity of the critical point of the 
binary liquid mixture, i.e., they are given as function of both the 
reduced temperature $t = (T - T_c) / T_c$ and the field $h$ 
conjugate to the order parameter (see Fig.\,\ref{fig_pdp} 
and Sec.\,\ref{sec_ses}).

In an ensemble of colloidal particles dissolved in a near-critical
solvent, the critical Casimir forces between them are expected to lead 
to flocculation of the particles. This holds 
even in cases where dispersion forces alone are not strong enough
to produce such a phase transition (see Subsec.\,\ref{sub_dis}). Indeed,
such a flocculation of colloidal particles has been observed
experimentally for various binary liquid mixtures 
\cite{BE,GKM92,LPB98,NK93,JK97,GW97}.
In particular, these experiments exhibit an asymmetry in the 
shape of the observed flocculation diagrams in that flocculation
occurs on that side of the phase diagram where
the binary liquid mixture is poor in the component 
preferred by the colloids. Our results are consistent
with this observation, since we indeed find that the critical 
Casimir forces are strongest in this region of the phase diagram
(see Subsecs.\,\ref{sub_hsn}, \ref{sub_grn}, and \ref{sub_sre}).

\bigskip
In the following we summarize the main results of this work:
\begin{enumerate}
\item{
For two spherical particles of radius $R$ at a distance $D$ 
(see Fig.$\,$\ref{fig_geo}) we have investigated the effective interaction 
mediated by the solvent, such as a binary liquid mixture. This effective
interaction depends on the thermodynamic state of the solvent, for
which the phase diagram is sketched in Fig.$\,$\ref{fig_pdp}.
}
\item{
The solvent is described by a field-theoretic model, provided by the 
Ginzburg-Landau Hamiltonian [Eq.$\,$(\ref{I100})],
with additional terms representing
the solute spheres. We solve this model within mean-field theory as 
described in Sec. \ref{sec_num}.
}
\item{
(\ref{sub_con}) The profile at the critical point is obtained by a
conformal mapping of the corresponding profiles between concentric
spheres. (\ref{sub_sde}) Close to the surface of either sphere the
{\em short distance expansion} of the profile is valid, which is
carried out off the critical point. (\ref{sub_min}) For the determination
of the complete profiles we first discretize the space using the
{\em conformal lattice} shown in Fig.$\,$\ref{fig_cnfl} and then apply
the method of steepest descent for the order parameter at the lattice points.
(\ref{sub_st}) From the order parameter profiles the force between the
spheres is calculated
by integrating the {\em stress tensor}. The choice of this integration 
path is indicated in Fig.$\,$\ref{fig_sts}.
}
\item{
The results for the two-particle interaction are presented in 
Sec.$\,$\ref{sec_ses}. (\ref{sub_tsn}) The dependence of the force on the
temperature at the critical composition is shown in Fig.$\,$\ref{fig_tscn}.
It exhibits a maximum above $T_c$ which approaches that of the critical point
for large distances. (\ref{sub_hsn}) The dependence of the force on the
composition at the critical temperature is shown in Fig.$\,$\ref{fig_hscn}.
It exhibits a pronounced maximum whose position also approaches 
that of the critical point
for large distances. The values of the maximum of the force determined
for various distances are shown in Fig.$\,$\ref{fig_maxf}. For small 
distances these values differ significantly from the value at the critical 
point. (\ref{sub_grn}) In Fig.$\,$\ref{fig_escn} the dependence of the force
on the temperature for fixed compositions off the critical one is shown.
(\ref{sub_der}) Using the knowledge of the force between parallel plates
the {\em Derjaguin approximation} can be applied for small particle
distances. (\ref{sub_sre}) The {\em small radius expansion} can be applied
for large particle distances regarding the spheres as perturbation
of the bulk phase. This expansion yields nonperturbative results for $d=3$.
In leading order the small radius expansion leads to
forces which exhibit a maximum only at the critical point.
(\ref{sub_dis}) In Fig.$\,$\ref{fig_frl} a comparison of the Casimir forces 
and the dispersion forces at $T_c$ and in $d=3$ is shown.
For large distances the
dispersion forces decay faster while for small distances the amplitudes
are smaller than those for the Casimir forces. Moreover, the temperature
and composition dependences of the Casimir forces allow one to distinguish them
from the dispersion forces. (\ref{sub_eos}) 
Relations between thermodynamic quantities as they are used in the present
model and their experimental counterparts are discussed.
}
\item{
In Sec.$\,$\ref{sec_flm} results for the case of parallel plates are presented,
which are used via the Derjaguin approximation for the limiting case of
small particle distances. Moreover, the present mean-field results are
in qualitative agreement with corresponding results for a two-dimensional
Ising spin system obtained in ref. {}\cite{DME00}. Figure$\,$\ref{fig_kvh}
describes the dependence of the force on the bulk field $H$ at the 
critical temperature.
In Fig.$\,$\ref{fig_cc} the dependence of the force on the 
bulk field is shown also for temperatures considerably below $T_c$. 
As $H$ is increased the force jumps from a small absolute value to a large one.
This is accompanied by a first-order phase transition of the corresponding
order parameter profiles as shown in Fig.$\,$\ref{fig_trns}. These first-order
phase transitions correspond to capillary condensation in the slit and their
loci are displayed in Fig.$\,$\ref{fig_ccpd}. The transition line ends in the
capillary critical point. In Fig.$\,$\ref{fig_df} the size of the jump
of the force at capillary condensation (see Fig.$\,$\ref{fig_cc})
is plotted as function of the bulk field.
Away from the capillary critical point this difference is
well approximated by an estimate in which only the order parameter value
for $H=0$ at the transition temperature
and the corresponding bulk field enter.
}
\end{enumerate}

%%%%%%%%%%%%%%%%%%%%%%%%%%%%%%%%%%%%%%%%%%%%%%%%%%%%%%%%%%%%%%%%%%
\section*{Acknowledgments}
We thank E. Eisenriegler and A. Macio{\l}ek for helpful and
stimulating discussions. We acknowledge partial financial support by the
German Science Foundation through Sonderforschungsbereich 237 (FS)
and grant No. HA3030/1-2 (AH). AH also acknowledges financial support by the 
National Science Foundation through grant 6892372 and by the Engineering and 
Physical Sciences Research Council through grant GR/J78327. 

%%%%%%%%%%%%%%%%%%%%%%%%%%%%%%%%%%%%%%%%%%%%%%%%%%%%%%%%%%%%%%%%%%

%%%%%%%%%%%%%%%%%%%%%%%%%%%%%%%%%%%%%%%%%%%%%%%%%%%%%%%%%%%%%%%%%%

\addtocounter{figure}{+1}
\begin{figure}[h]
\setlength{\epsfxsize}{12cm}
\centerline{\epsfbox{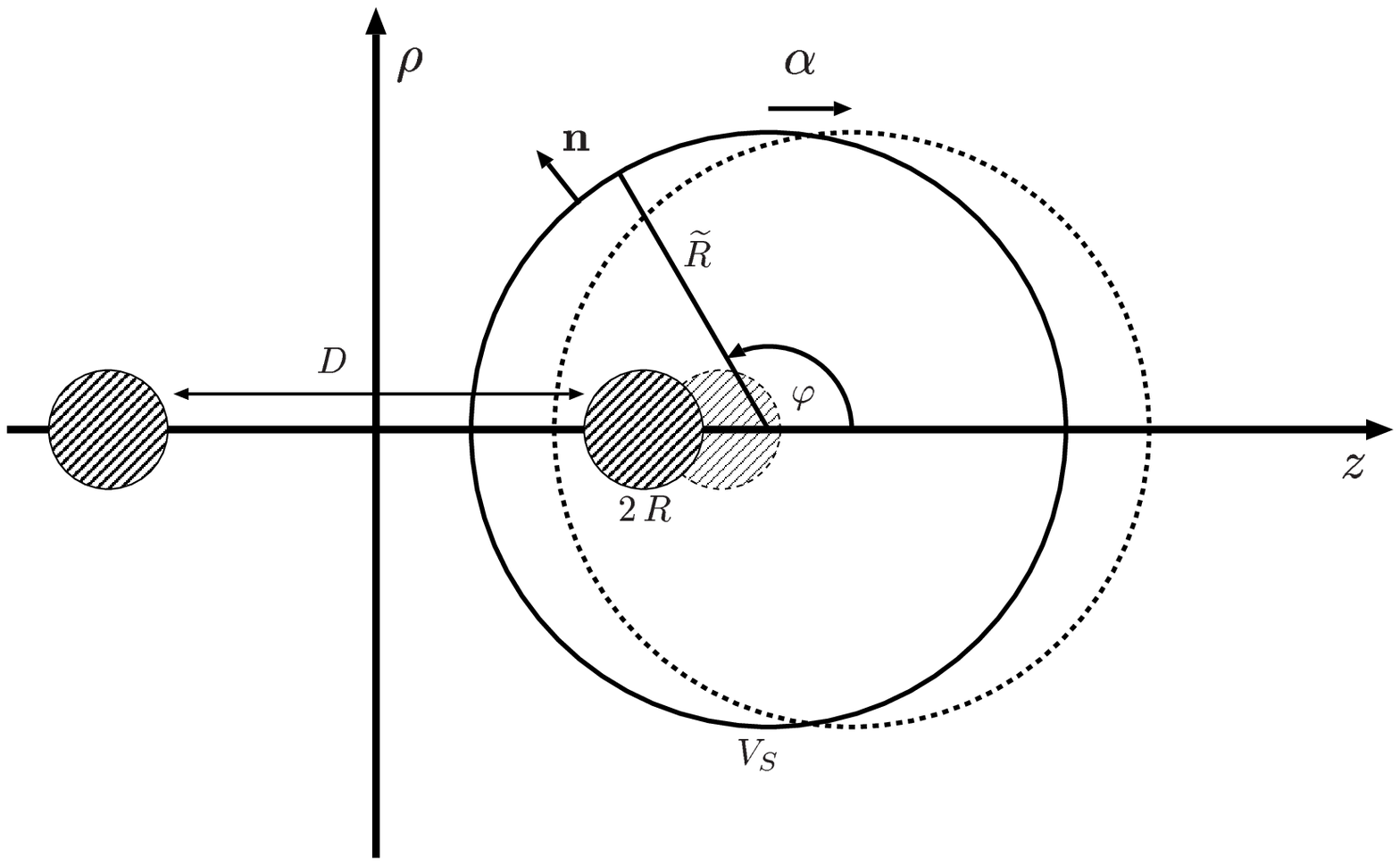}}
\caption{\label{fig_sts} Calculating the force via Eq.$\:$(\ref{KglInt}).
$\widetilde{R}$ is the radius of a sphere $V_S$ that contains one of
the spherical particles of diameter $2\,R$. The projection onto the 
$\rho$-$z$-plane is parametrized
with the angle $\varphi$. The interior of $V_S$ is shifted by the 
infinitesimal amount $\alpha$, which is equivalent to an infinitesimal
change of the distance $D$ between the two particles.}
\end{figure}
\addtocounter{figure}{-2}
\begin{figure}[p]
\setlength{\epsfxsize}{15cm}
\centerline{\epsfbox{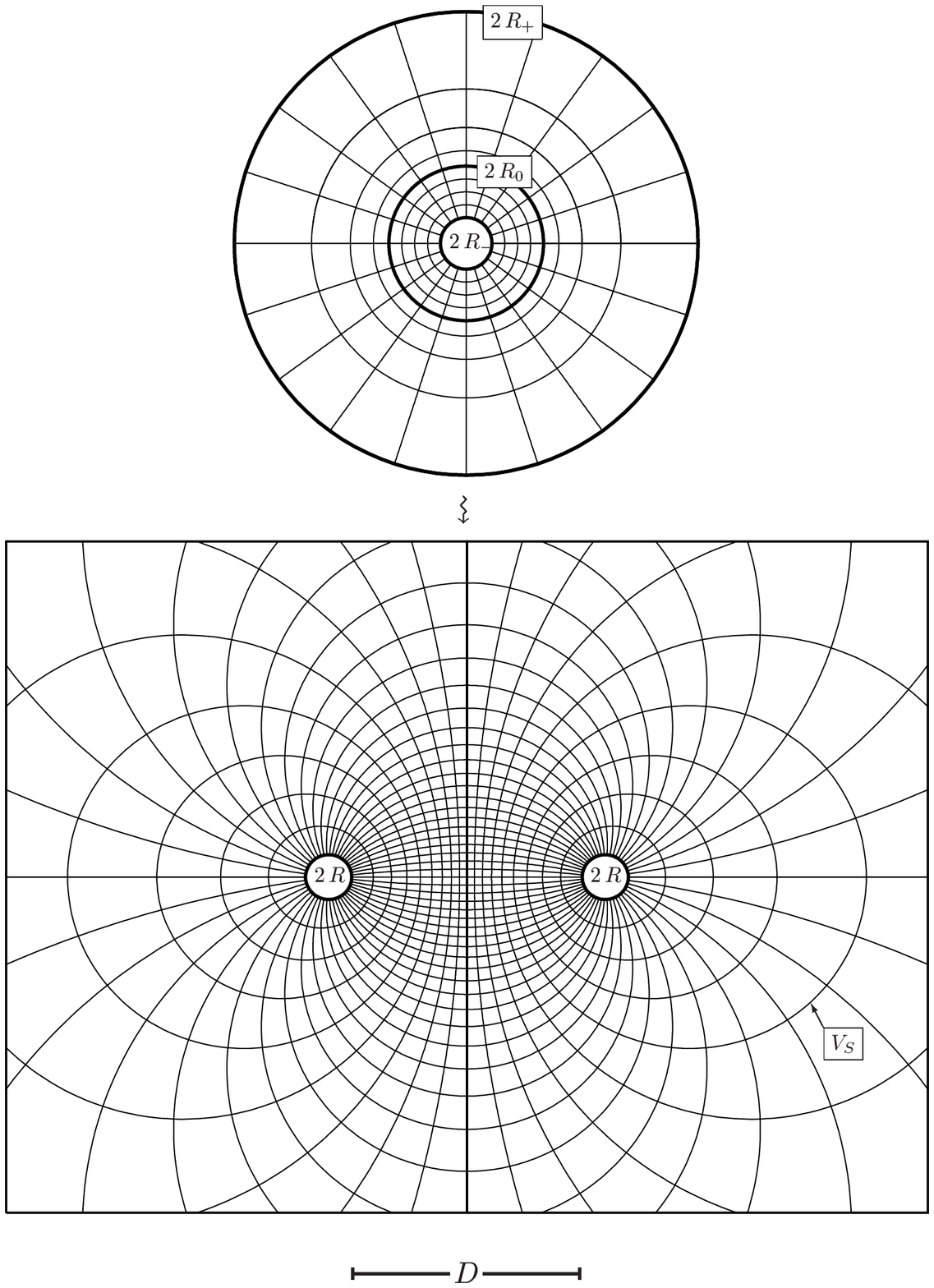}}
\caption{\label{fig_cnfl} The conformal lattice (bottom)
for two spheres of diameter $2\,R$ at distance $D$ (see Fig.$\,$\ref{fig_geo})
as obtained from the lattice between concentric spheres of diameter $2\,R_+$ 
and $2\,R_-$, respectively. (For the purpose of clear visibility here the two
lattices are shown on different scales.)
The sphere with diameter $2\,R_0=2\,\sqrt{R_+\,R_-}$ is mapped onto
the vertical midplane of the conformal lattice.
We note that inside the sphere with diameter $2\,R_0$ the spherical lattice
lines are chosen equidistant and more dense than outside.
$R$ and $D$ are given by Eqs.$\,$(\ref{IRpm1}) and 
(\ref{IRpm3}).
$V_S$ is introduced in Subsec.$\,$\ref{sub_st}.
}
\end{figure}
\addtocounter{figure}{+1}
\begin{figure}[h]
\setlength{\epsfxsize}{11cm}
\centerline{\epsfbox{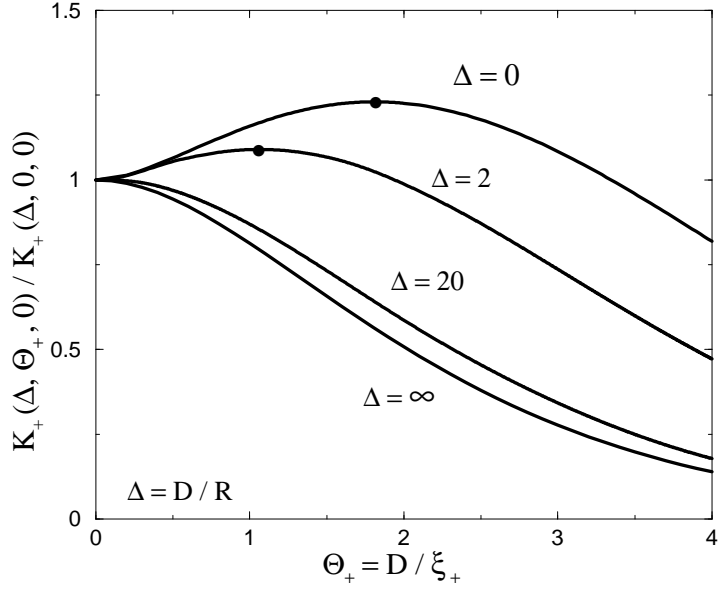}}
\caption{\label{fig_tscn} Force between two particles as function of 
temperature $T>T_c$ and for $H=0$
relative to the force at criticality. The force is maximal for
$\Theta_+=\Theta_{\rm max}(\Delta)$ as indicated by the dots.}
\end{figure}
\begin{figure}[h]
\setlength{\epsfxsize}{11cm}
\centerline{\epsfbox{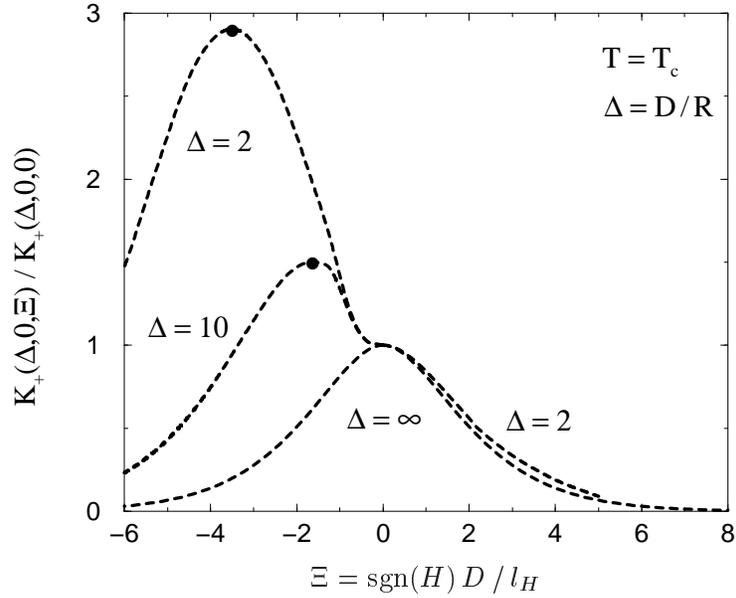}}
\caption{\label{fig_hscn} Force between two particles as function of 
the bulk field $H$ at $T=T_c$ relative to the force at criticality 
$(\tau,H)=(0,0)$. The force is maximal for
$\Xi=\Xi_{\rm max}(\Delta)$ as indicated by the dots.}
\end{figure}
\begin{figure}[h]
\setlength{\epsfxsize}{10cm}
\centerline{\epsfbox{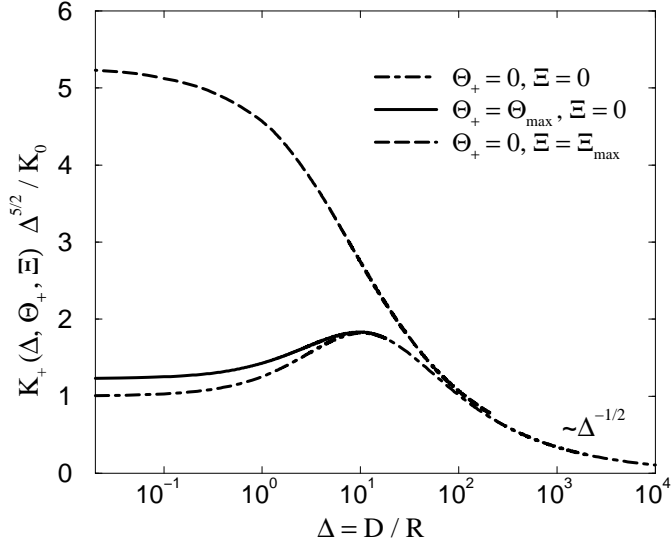}}
\caption{\label{fig_maxf} Maximum force values as function of distance.
For small distances the force increases as $\Delta^{-5/2}$ which is scaled
out here. Accordingly, the forces are normalized by the value 
$K_0=\lim_{\Delta\to 0}(\Delta^{5/2}K_{+}(\Delta,0,0))$ at the critical point 
in the limit of vanishing distance. For large distances the forces decay
as $\Delta^{-2\,\beta/\nu-1}$, i.e., as $\Delta^{-3}$
within mean-field theory so that $\Delta^{5/2}K_{+}$ decays as 
$\Delta^{-1/2}$. $\Theta_+=D/\xi_+$ and $\Xi=\text{sgn}(H)D/l_H$.}
\end{figure}
\begin{figure}[h]
\setlength{\epsfxsize}{10cm}
\centerline{\epsfbox{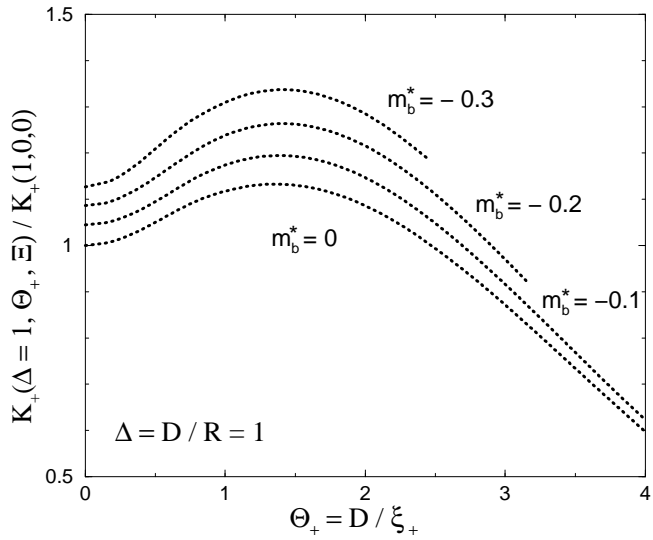}}
\caption{\label{fig_escn} Force between two particles as function of 
temperature for fixed compositions $c_A\neq c_{A,c}$
which implies fixed values of ${m_b}^{\ast}=m_b\,D^{\beta/\nu}$. 
Considering the dotted path in Fig.$\,$\ref{fig_pdp}, i.e., for 
$m_b=\text{const}$ [for a relation between $m_b$ and $M_b\sim(c_A-c_{A,c})$
see, c.f., Eq.$\,$(\ref{mbMb})] both thermodynamic variables 
$\tau$ and $H$ and thus
both scaling variables $\Theta_+=D/\xi_+$ and $\Xi=\text{sgn}(H)D/l_H$
are varied simultaneously [see, c.f., Eq.$\,$\ref{eosmf}].
The Casimir force increases by going further away from the critical 
composition.
}
\end{figure}
\begin{figure}[h]
\setlength{\epsfxsize}{10cm}
\centerline{\epsfbox{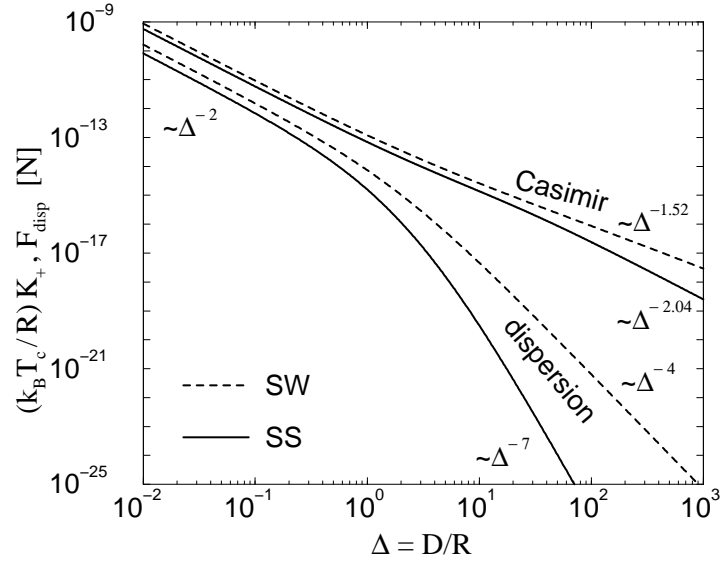}}
\caption{\label{fig_frl} Comparison of dispersion forces and Casimir forces
at the critical point ($T=T_c, c_A=c_{A,c}$)
as function of the distance $D$ in units of the sphere radius $R$. The forces 
are shown
for both a sphere in front of a planar wall (SW) and for two spherical
particles (SS). For a detailed discussion see the main text.}
\end{figure}
\begin{figure}[h]
\setlength{\epsfxsize}{10cm}
\centerline{\epsfbox{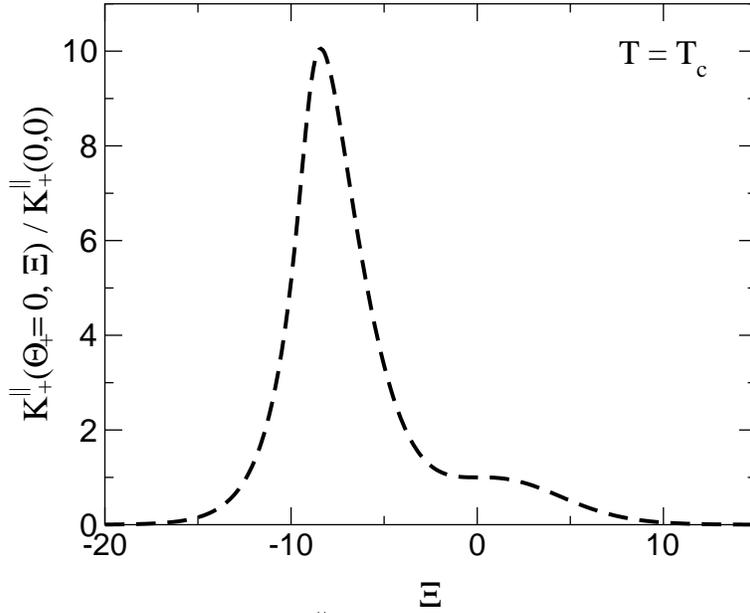}}
\caption{\label{fig_kvh}Normalized scaling function $K_{+}^{||}$ of the 
force for a film with thickness $L$ at $T_c$, i.e., $\Theta_+=L/\xi_+=0$ as 
function of the scaling variable $\Xi=\text{sgn}(H)\,L/l_H$.}
\end{figure}
\begin{figure}[h]
\setlength{\epsfxsize}{11cm}
\centerline{\epsfbox{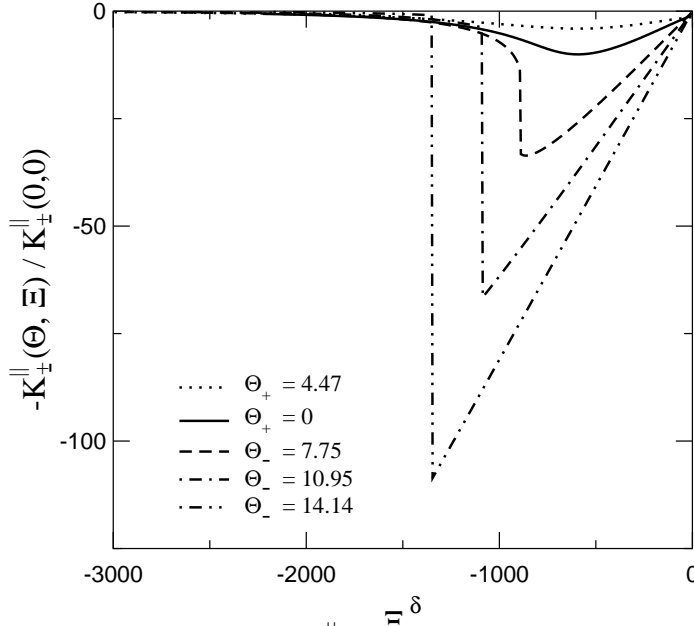}}
\caption{\label{fig_cc}Normalized scaling function $K_{\pm}^{||}$ 
of the force between parallel plates at distance $L$ for different 
temperatures, i.e., different values of the scaling variable
$\Theta_{\pm}=L/\xi_{\pm}$
as function of the bulk field $H$. Note that the scaling
variable $\Xi=\text{sgn}(H)\,L/l_H$ raised to the power $\delta$,
where $\delta=3$ within mean-field theory, is directly proportional 
to $H$ revealing the almost linear dependence of the force on $H$
over a wide range of values.
For details see the main text.}
\end{figure}
\begin{figure}[h]
\setlength{\epsfxsize}{9cm}
\centerline{\epsfbox{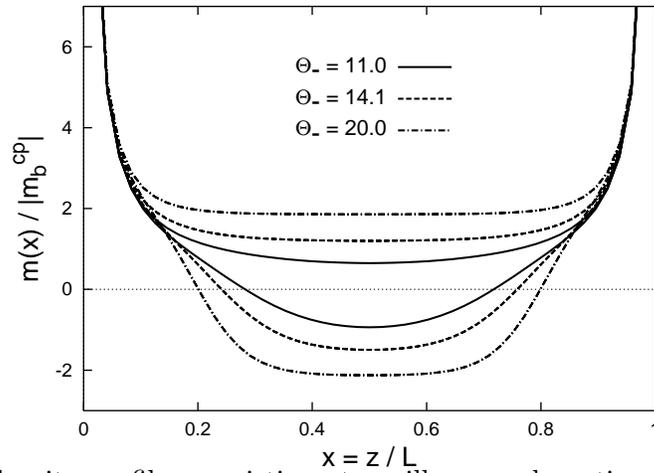}}
\caption{\label{fig_trns}Pairs of
density profiles coexisting at capillary condensation for various
temperatures $T<T_c$ relative to the bulk order parameter $m_b^{cp}$
at the capillary critical point. The lower curves are stable further 
away from bulk coexistence whereas for the upper ones capillary condensation 
has taken place.
Above the critical point of capillary condensation (see Fig.$\,$\ref{fig_ccpd})
the profiles look similar to the upper ones.}
\end{figure}
\begin{figure}[h]
\setlength{\epsfxsize}{9cm}
\centerline{\epsfbox{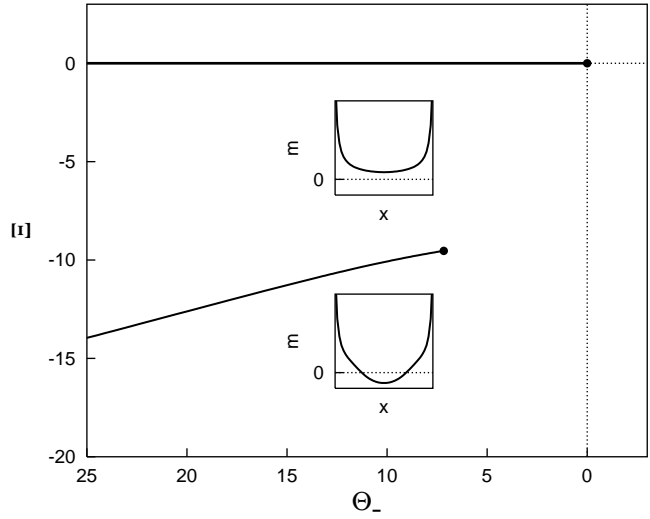}}
\caption{\label{fig_ccpd}Phase diagram for capillary condensation as obtained
within the present mean field approach. The transition line ends in
the capillary critical point ($\Theta_-=7.17,\Xi=-9.54$).
The straight line ending at ($\Theta_-=0,\Xi=0$) is the line of first-order 
phase transitions in the bulk.
}
\end{figure}
\begin{figure}[h]
\setlength{\epsfxsize}{9cm}
\centerline{\epsfbox{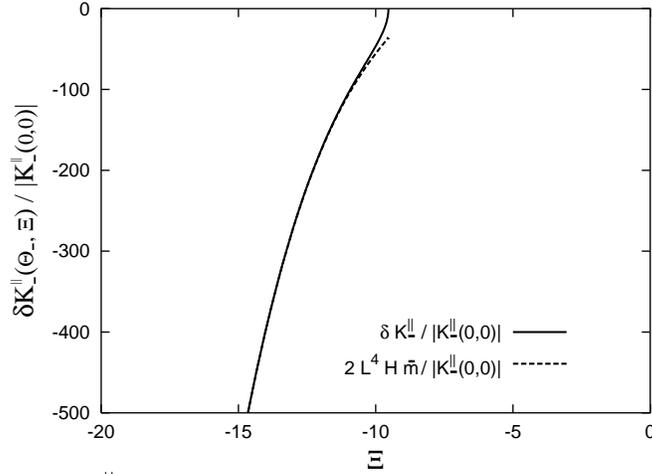}}
\caption{\label{fig_df}The difference $\delta K_{-}^{||}$ between the force 
values of profiles coexisting at capillary condensation (solid line) and an 
approximate expression thereof (dashed line). 
The overall $L$ dependence of the forces ($L^{-4}$) is scaled out
in $K_{-}^{||}$, $\delta K_{-}^{||}$, and the approximation.}
\end{figure}

\end{document}